\documentstyle[11pt,a4,epsfig,amssymb]{article}

\makeatletter
\@addtoreset{equation}{section}
\makeatother

\newcommand{\mrm}[1]{\mathrm{#1}}
\newcommand{\mbf}[1]{\mathbf{#1}}

\title{
\begin{flushright} {\normalsize LC-PHSM-2006-002} \end{flushright}
\vspace{6ex}
{\bf
Measurement of the top-Higgs Yukawa coupling at a Linear $e^{+}e^{-}$ Collider} \\
\vspace{2ex}
\vspace{1ex}
}

\date{}

\author{
\large {\sc Arnaud GAY \footnote{e-mail: arnaud.gay@ires.in2p3.fr}} \\
\\
\normalsize Institut Pluridisciplinaire Hubert Curien\\
23 rue du Loess, BP 20, F-67037 Strasbourg, France
}

\begin{document}

\maketitle

\begin{abstract}

Understanding the mechanism of electroweak symmetry breaking and the origin of boson and fermion masses is among the most pressing questions raised in contemporary particle physics. If these issues involve one (several) Higgs boson(s), a precise measurement of all its (their) properties will be of prime importance. Among those, the Higgs coupling to matter fermions (the Yukawa coupling). At a Linear Collider, the process $e^{+}e^{-} \rightarrow t\overline tH$ will allow in principle a direct measurement of the top-Higgs Yukawa coupling. We present a realistic feasibility study of the measurement in the context of the TESLA collider. Four channels are studied and the analysis is repeated for several Higgs mass values within the range 120 GeV/c$^{2}$ - 200 GeV/c$^{2}$.

\end{abstract}

\newpage
\tableofcontents

\newpage

\section{Introduction}

The gauge sector of electroweak interactions has been checked to coincide with the Standard Model (SM) prediction to the per-mil level, at LEP and SLC. On the contrary, there is no direct experimental evidence for the Higgs mechanism, supposed to be responsible for electroweak symmetry breaking and the generation of masses. Direct search of the Higgs boson at LEP yields the lower limit \cite{lephiggs}: $M_{H} > 114.4$ GeV/c$^{2}$ at $95\%$ CL. Precision measurements on the other hand give \cite{lepeww}: $M_{H} \lesssim 240$ GeV/c$^{2}$ at $99\%$ CL. Once a Higgs particle is found, if ever, all its properties should be measured precisely to completely characterise the Higgs mechanism. Among those, the coupling of the Higgs boson to fermions (the Yukawa coupling), which is supposed to scale with the fermion mass:

\begin{equation}
g_{ffH} = \frac{m_{f}}{v}
\end{equation}

\noindent where $g_{ffH}$ is the Yukawa coupling of a fermion f of mass $m_{f}$ and $v$ is the vacuum expectation value of the Higgs field, $v=({\sqrt{2}G_{F}})^{-1/2} \approx 246$ GeV.

The top quark is the heaviest fermion, thus the top-Higgs Yukawa coupling should be the easiest to measure. If $M_{H} >2*m_{t}$, this parameter can be measured through the branching ratio of the Higgs boson decay into a pair of top quarks. Otherwise, i.e. for lower values of the Higgs boson mass, the process $e^{+}e^{-} \rightarrow t\overline tH$ allows in principle a direct measurement of this coupling.

Feasibility studies of the measurement of the top-Higgs Yukawa coupling via the process $e^{+}e^{-} \rightarrow t\overline tH$ at a Linear Collider have already been performed ~\cite{yukawaseq}~\cite{justemerino} for a Higgs boson mass of 120-130 GeV/c$^{2}$. This is the most favourable case (taking into account the lower mass bound) as the cross-section of this process decreases with increasing Higgs boson mass and as a Higgs boson of such a mass decays predominantly to a pair of b quarks, allowing a very effective signal and background separation using b-tagging algorithms. One of the studies (\cite{justemerino}) showed that a neural network analysis was essential to get a precise result. We repeated this work and extended it up to $M_{H} =$ 150 GeV/c$^{2}$. When $M_{H} \gtrsim $ 135 GeV/c$^{2}$, the $H \rightarrow W^{+}W^{-}$ decay mode dominates. This channel was also studied, for masses up to 200 GeV/c$^{2}$.

\section{The process $e^{+}e^{-} \rightarrow t\overline tH$}

\begin{figure}[h]
\begin{center}
\psfig{figure=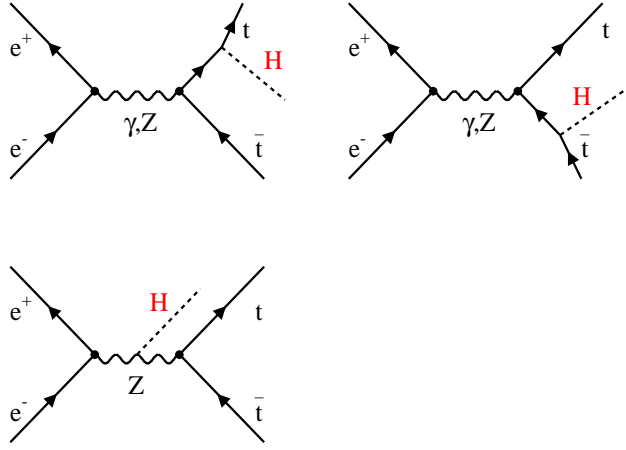,height=7cm,width=9cm}
\end{center}
\caption{\it Lowest order Feynman diagrams of the process $e^{+}e^{-} \rightarrow t\overline tH$.}
\label{diagrammetth} 
\end{figure}

The lowest order Feynman diagrams contributing to the $e^{+}e^{-} \rightarrow t\overline tH$ process are shown in figure~\ref{diagrammetth}. The amplitude of the diagram where the Higgs boson is radiated from the Z boson is not expressing the top-Higgs Yukawa coupling. However, since it modifies only slightly the cross-section of the process, it can safely be neglected. The cross-section and the top-Higgs Yukawa coupling thus verify to a good approximation: $ \sigma _{e^{+}e^{-} \rightarrow t\overline tH} \propto g^{2}_{ttH} $.

\begin{figure}[h]
\begin{center}
\psfig{figure= 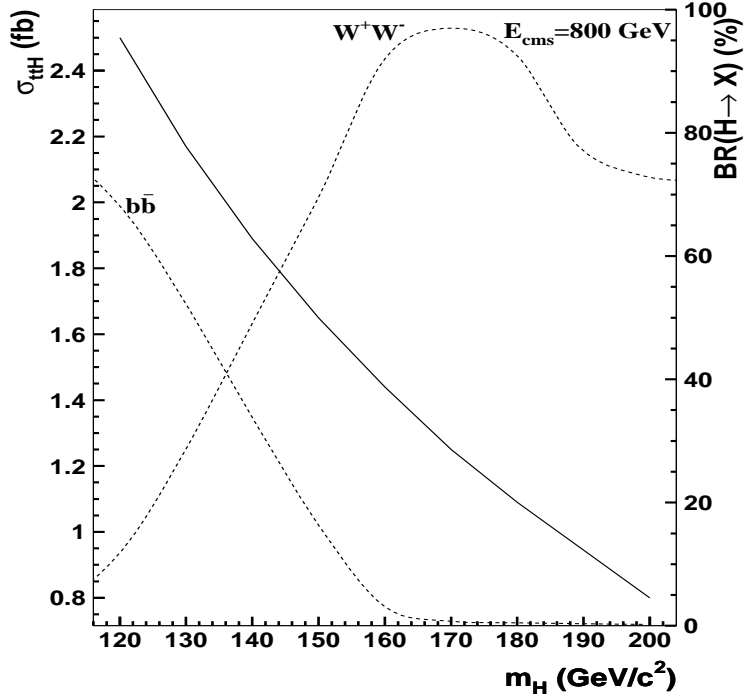,height=10cm,width=10cm}
\end{center}
\caption{\it Higgs branching ratios (dashed lines) for the $H \rightarrow b \overline b$  and $H \rightarrow W^{+}W^{-}$ modes (as given by HDECAY) and cross-section (solid line) at lowest order of the process $e^{+}e^{-} \rightarrow t\overline tH$ (as given by CompHEP), for various Higgs mass values and for $\sqrt s =$ 800 GeV. In the calculation of the cross-section, initial state radiation and beamstrahlung were taken into account.}
\label{crosssectiontth} 
\end{figure}

For this work, the following assumptions were made: $m_{t}=175$ GeV/c$^{2}$ and $BR(t\rightarrow Wb)=100\%$. The Higgs branching ratios were calculated with the HDECAY~\cite{hdecay} program. The values obtained for the $H \rightarrow b \overline b$  and $H \rightarrow W^{+}W^{-}$ modes, which are the main decays within the Higgs mass range considered in this paper, are shown in table~\ref{brhiggs} and figure~\ref{crosssectiontth}.

\begin{table}[h!]
\centering
\begin{tabular}{||c||c||c||c|}
\hline
\hline
{ $M_{H}$ (GeV/c$^{2}$)} & {$BR(H \rightarrow b\overline b)$}& {$BR(H \rightarrow W^{+}W^{-})$} & {$\sigma$ (fb)}\\
{ } & { } & { } & {($\sqrt s =$ 800 GeV) } \\ 
\hline
{ 120} & {67$\%$}& {13$\%$}& {2.50}\\
{ 130} & {51$\%$}& {30$\%$}& {2.17}\\
{ 140} & {33$\%$}& {50$\%$}& {1.88}\\
{ 150} & {16$\%$}& {70$\%$}& {1.64}\\
{ 160} & {$3.1\%$}& {$92\%$}& {1.44}\\
{ 170} & {$0.76\%$}& {$97\%$}& {1.25}\\
{ 180} & {$0.48\%$}& {$93\%$}& {1.09}\\
{ 200} & {$0.23\%$}& {$73\%$}& {0.80}\\
\hline
\hline
\end{tabular}
\caption{\it Higgs branching ratios for the $H \rightarrow b \overline b$  and $H \rightarrow W^{+}W^{-}$ modes (as given by HDECAY) and cross-section at lowest order of the process $e^{+}e^{-} \rightarrow t\overline tH$ (as given by CompHEP), for various Higgs mass values and for $\sqrt s =$ 800 GeV. In the calculation of the cross-section, initial state radiation and beamstrahlung were taken into account.}
\label{brhiggs}
\end{table}

Previous studies showed that a center-of-mass energy close to the kinematical limit of the $t\overline tH$ production process does not allow to measure the coupling with a satisfactory precision, due to the tiny cross-section. Center-of-mass energies of 800 GeV and 1 TeV have shown to be far better for the measurement. A run at 800 GeV being a very likely possibility for the Linear Collider, this will be the default value of our study. The lowest order cross-section of the process (as given by the CompHEP~\cite{comphep} software)  is shown in table~\ref{brhiggs} and figure~\ref{crosssectiontth} for the Higgs mass range considered and a center-of-mass energy of 800 GeV.

The ${\cal O}(\alpha_s)$ corrections to the $e^{+}e^{-} \rightarrow t\overline tH$ process have been calculated by several groups~\cite{tthcor}. At a center-of-mass energy of 800 GeV, they affect the total cross-section by less than $5\%$ and were thus neglected in this study.

\section{Measurement of $g_{ttH}$}

For a particular analysis yielding a selection efficiency of the signal $\epsilon_{sel}^{signal}$  and a purity of the selected sample $\rho_{sel}^{sample}$ and assuming an integrated luminosity L, the statistical and systematic uncertainties on the measurement of $g_{ttH}$ can be expressed as follows:

\begin{equation}
\big(\frac{\Delta g_{ttH}}{g_{ttH}}\big)_{stat} \approx \frac{1}{S_{stat}(g^{2}_{ttH}) \sqrt{ \epsilon_{sel}^{signal} \rho_{sel}^{sample} L} }
\end{equation}

\begin{equation}
\big(\frac{\Delta g_{ttH}}{g_{ttH}}\big)_{syst} \approx \frac{1}{S_{syst}(g^{2}_{ttH})  }\frac{1-\rho_{sel}^{sample}}{\rho_{sel}^{sample}}\frac{\Delta \sigma _{BG}^{eff}}{\sigma _{BG}^{eff}}
\end{equation}

The default value for L assumed through the whole study is 1000 $fb^{-1}$. This large value is quite essential to maintain the statistical uncertainty at the level of a few per-cent. $\frac{\Delta \sigma _{BG}^{eff}}{\sigma _{BG}^{eff}}$ is the relative uncertainty on the residual background normalisation. It is mostly due to badly known differential cross-sections in weakly populated phase space corners. It is sizeable and moreover difficult to estimate. However, $t\overline t$ pairs will be copiously produced at the LC, allowing a complete characterisation of them. For example, studying these pairs during a run at an energy where the $t\overline tH$ process is negligible will allow to improve their simulation by event generators. Therefore, this uncertainty should not exceed 10\%. We will thus repeat the analysis for two values of this uncertainty, $10\%$ and $5\%$. In the systematic uncertainty, we just take into account the one which arises from the effective background normalisation since it is by far the largest one among those we can estimate now.

The sensitivity factors $S_{stat}$ and $S_{syst}$ in relations 3.1 and 3.2 express the dependence of the cross-section on the coupling squared:

\begin{equation}
S_{stat}(g^{2}_{ttH}) =\frac{1}{\sqrt{\sigma _{t\overline tH}}} \Bigg\vert  \frac{d\sigma _{t\overline tH}}{dg^{2}_{ttH}} \Bigg\vert
\end{equation}

\begin{equation}
S_{syst}(g^{2}_{ttH}) = \frac{1}{\sigma _{t\overline tH}} \Bigg\vert \frac{d\sigma _{t\overline tH}}{dg^{2}_{ttH}} \Bigg\vert
\end{equation}

As pointed out in section 2, the contribution from Higgs radiation off the Z to the signal cross-section is very small. In order to calculate the sensitivity factors, we will thus neglect it, allowing a very simple calculation. In this approximation, we can write:

\begin{equation}\label{eq1}
\sigma_{t\overline tH} \approx g^{2}_{ttH} F(M_{H},m_{t},s)
\end{equation}

\noindent And thus:
\begin{equation}\label{eq2}
\frac{d\sigma _{t\overline tH}}{dg^{2}_{ttH}} \approx F(M_{H},m_{t},s) \approx \frac{\sigma _{t\overline tH}}{g^{2}_{ttH}}
\end{equation}

\noindent where s is the squared collision energy. 

\noindent The sensitivity factors eventually read:

\begin{equation}
S_{stat}(g^{2}_{ttH}) =\frac{\sqrt{\sigma _{t\overline tH}}}{g^{2}_{ttH}}
\end{equation}

\begin{equation}
S_{syst}(g^{2}_{ttH}) = \frac{1}{g^{2}_{ttH}}
\end{equation}

The values of $S_{stat}$ and $S_{syst}$ are summarised in table~\ref{Sensifac}. In this approximation, $S_{syst}$ is independent of the Higgs mass and of the center-of-mass energy. The results obtained for $M_{H}=120$ GeV/c$^{2}$ agree with those presented in~\cite{justemerino}.

\begin{table}[h]
\centering
\begin{tabular}{||c||c||c||}
\hline
\hline
{ $M_{H}$ (GeV/c$^{2}$)} & {$S_{stat}$ (fb$^{1/2}$)} & {$S_{syst}$}\\
\hline
{ 120} & {3.13} & {1.98} \\
{ 130} & {2.92} & {1.98} \\
{ 140} & {2.72} & {1.98}\\
{ 150} & {2.54} & {1.98}\\
{ 170} & {2.21} & {1.98}\\
{ 200} & {1.77} & {1.98}\\
\hline
\hline
\end{tabular}
\caption{\it Sensitivity factors for various Higgs mass values at $\sqrt s =$ 800 GeV.}
\label{Sensifac}
\end{table}

\section{Analysis and simulation}
\subsection{Background}

The resonant background processes considered in the analysis are listed in table~\ref{sectionefficacebg}, together with their cross-sections (as given by the CompHEP software).

\begin{table}[h]
\centering
\begin{tabular}{||c||c||}
\hline
\hline
{ Elementary process} & {$\sigma$ (fb)} \\
\hline
{ $e^{+}e^{-} \rightarrow q\overline q$} & {1557.7} \\
{ $e^{+}e^{-} \rightarrow t\overline t$} & {297.3} \\
{ $e^{+}e^{-} \rightarrow W^{+}W^{-}$} & {4298} \\
{ $e^{+}e^{-} \rightarrow ZZ$} & {239.8} \\
{ $e^{+}e^{-} \rightarrow t\overline tZ$} & {4.3} \\
\hline
\hline
\end{tabular}
\caption{\it Resonant background cross-sections for $\sqrt s =$ 800 GeV from CompHEP. Initial state radiation and beamstrahlung are taken into account.}
\label{sectionefficacebg}
\end{table}

The processes $e^{+}e^{-} \rightarrow q\overline q$ and  $e^{+}e^{-} \rightarrow W^{+}W^{-}$ exhibit a very different topology from the signal, however, their huge cross-section (two or three orders of magnitude larger than the cross-section of the signal) forbids to neglect them. 

Although its cross section is much lower, the above statements hold for the process $e^{+}e^{-} \rightarrow ZZ$.

The process $e^{+}e^{-} \rightarrow t\overline tZ$ and its cross section are very close to those of the signal.

Finally, the process $e^{+}e^{-} \rightarrow t\overline t$  has a rather large cross-section and will often mimic the signal. It is expected to be the main background.

The backgrounds listed in table \ref{sectionefficacebg} don't account for 6 fermion production, which may turn out to be not negligible. The generation of these events is however a complicated and very time consuming task, which leads to rather large uncertainties. As we do not want to completely neglect these final states, we adopt the following procedure, which acts as a compromise. The selection is first optimised for the suppression of resonant background (which is by far the main background). The residual contamination from the dominant 6 fermion processes is then estimated with the same selection criteria. The results will be presented with and without inclusion of the 6 fermion processes in the overall background.

\subsection{Generation of events and simulation of the detector}

The $t\overline tH$ and $t\overline tZ$ partonic events were generated with CompHEP V.41.10. This program allows to include the initial state radiation and the beamstrahlung. These events were then treated with PYTHIA V.6.158~\cite{pythia} for hadronization, decay and final state radiation.

Resonant backgrounds were generated with PYTHIA V.6.158. In this case, the initial state radiation was considered in the structure function approach and the beamstrahlung was implemented with CIRCE~\cite{Ohl:1996fi}.

The 6 fermion process partonic events were generated with WHIZARD V.1.2x\footnote{Versions 1.22 to 1.24 were used.} \cite{whizard}. As for $t\overline tH$ and $t\overline tZ$ events, they were treated with PYTHIA V.6.158 for hadronization, decay and final state radiation. Beamstrahlung and initial state radiation are handled by WHIZARD.

Some of these 6 fermion processes receive contributions from the reactions listed in table 3. For these processes, the corresponding diagrams are removed from the calculation to avoid double counting, keeping however the interference between the resonant and non-resonant diagrams. This short-cut is motivated by the complexity a more correct treatment would introduce; it is justified by the magnitude of the interferences considered.

The cross-sections of the 6 fermion processes are in general calculated with quite good accuracy. However, when coming to the event generation, things get substantially more complicated and time consuming. Due to CPU limitations, the number of events in the 6 fermion samples is sometimes modest and, conversely, the differential cross sections have sizeable statistical uncertainties. Furthermore, some ambiguities may arise when pairing the particles in the final state (needed for hadronization and final state radiation) e.g. when there are 2 pairs of identical particles.
 
The contaminations by the 6 fermion processes will thus not be estimated as precisely as the ones by the resonant backgrounds. The loss of resolution arising from these events should thus be taken as a rough estimate. This shoud not matter as the residual 6 fermion background is small.

\vspace{0.5cm}

The events were further processed by SIMDET V.4~\cite{simdet}, the fast simulation program of the TESLA~\cite{tesla} detector, in order to take into account detector and event reconstruction effects.

\subsection{b-tagging}

The b-tagging is an essential tool for this analysis, especially when assuming that the Higgs boson decays into a pair of b-quarks. b and c-tagging algorithms developed for LEP and SLC experiments were adapted to TESLA and their performances studied~\cite{hawkings} ~\cite{smxh}. Recently, these tools were made available with the fast simulation of the detector~\cite{kuhl}. Various algorithms are combined, including the SLD-ZVTOP vertex finder, in a neural network. One of the outputs is the b-probability of a jet.

\subsection{Definition of the variables used in the analysis}

\subsubsection{Fox-Wolfram moments}

The Fox-Wolfram moments $H_l$, $l = 0, 1, 2, \ldots$, are defined by
\begin{equation}
H_l = \sum_{i,j} \frac{ |\mbf{p}_i| \, |\mbf{p}_j| }{E_{\mrm{vis}}^2} \, P_l (\cos \theta_{ij}),
\end{equation}
where $\theta_{ij}$ is the opening angle between hadrons $i$ and $j$ and $E_{\mrm{vis}}$ the total visible energy of the event. The $P_l(x)$ are the Legendre polynomials. In this paper, the moments will be normalized to $H_0$, i.e. $H_{l0} = H_l / H_0$.

\subsubsection{Heavy and light jet masses}

The particles of an event are divided into 2 classes and the invariant mass of each class is calculated. We obtain the heavy and light jet masses when the assignment of particles to the classes is such that the quadratic sum of the invariant mass of the 2 classes is minimised. In events with resonances (e.g. $e^{+}e^{-} \rightarrow W^{+}W^{-}$), these quantities tend to peak at their invariant mass.

\subsubsection{Other variables}

$E^{jet}_{max(min)}$ is the energy of the most (least) energetic jet of an event. The minimum jet multiplicity is the number of objects (charged as well as neutral) in the jet which has the smallest multiplicity. $P^{jet}_{b}(i)$ is the probability of the $i^{th}$ jet to be a b-jet, the $P^{jet}_{b}(i)$ being sorted out in decreasing order, i.e. from the most b-like to the least b-like.

The other variables are self-explanatory.

\subsection{Neural network analysis}

A neural network is used to optimize the selection of events. We used the MLPfit program \cite{mlpfit}, a multi-layer perceptron with error backpropagation.

\section{Study of the $H \rightarrow b\overline b$ decay mode}

\subsection{Introduction}

When the Higgs boson decays into $ b\overline b$ pairs, 3 classes of final states occur. Among them, two can potentially allow to measure the Yukawa coupling well: the semileptonic final state ($t\overline t \rightarrow W^{+}bW^{-}\overline b \rightarrow 2b2ql\nu $ with $BR(t\overline t \rightarrow 2b2ql\nu) \approx 43.9\%) $ and the hadronic final state ($t\overline t \rightarrow W^{+}bW^{-}\overline b \rightarrow 2b4q$ with $BR(t\overline t \rightarrow 2b4q) \approx 45.6\%)$. These two channels are characterized by a large particle and jet multiplicity and an isotropic topology. The presence of four b-jets will allow the construction of very discriminating variables. However, the event rate is really tiny in comparison with the background and the very crowded environnement will degrade clustering and b-tagging algorithms and will make invariant mass constraints less effective. Furthermore, hard gluon radiation combined with gluon splitting to $b\overline b$ will often allow the background events to fake the signal.

For both channels, a preselection sequential procedure is first applied in order to remove most of the background, while keeping a high selection efficiency for the signal. A neural network analysis will then be performed in order to optimally use the information contained in the distributions of the final state characteristics. The analysis will be repeated for the following Higgs boson masses: 120 GeV/c$^{2}$, 130 GeV/c$^{2}$, 140 GeV/c$^{2}$ and 150 GeV/c$^{2}$. As the distributions hardly change in this mass window, the same procedure will be applied for each value of the Higgs boson mass.

\subsection{Semileptonic Channel}\label{semilep}
\subsubsection{Introduction}

The final state follows from the process:
\\

$e^{+}e^{-} \rightarrow t\overline tH \rightarrow W^{+}bW^{-}\overline b b\overline b \rightarrow 4b2ql\nu$.
\\

This channel is thus characterized by 4 b-jets, 2 light quark jets, one prompt lepton and missing 4-momentum. It has a little less statistics than the hadronic channel but the final state is cleaner and the presence of an isolated lepton together with missing 4-momentum will allow to construct powerful selection variables. Notice that, here, the hadronic channel ($t\overline t H \rightarrow 4b4q$), the fully leptonic channel ($t\overline t H \rightarrow 4b2l2\nu$) and the channels where the Higgs boson doesn't decay into b quarks act as background processes.

\subsubsection{Sequential analysis}

\begin{figure}
\begin{center}
\epsfig{figure=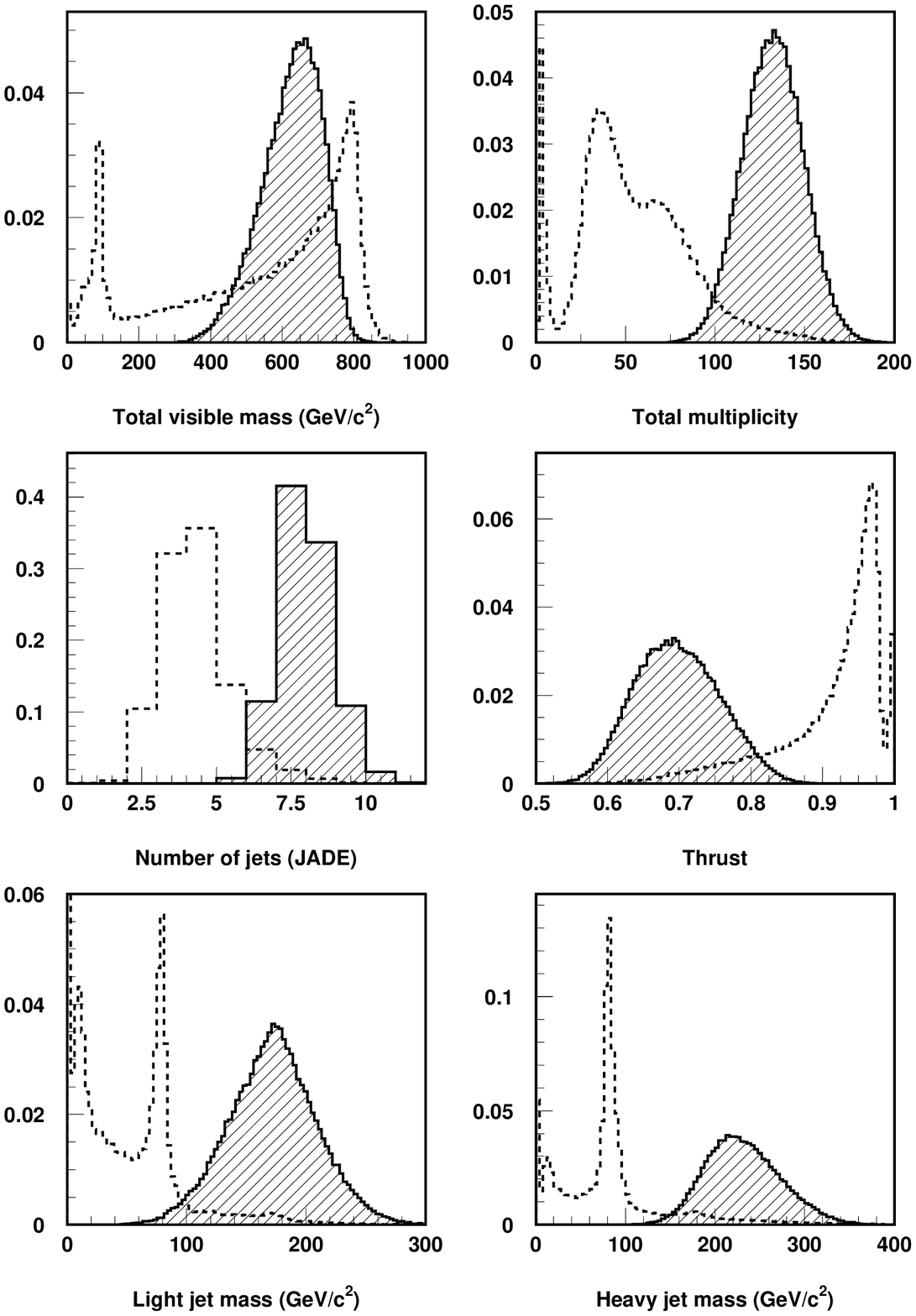,height=17cm,width=15cm}
\end{center}
\caption{\it The $H \rightarrow b \overline b$ semileptonic channel: preselection variables (I). The signal (solid line) and the background (dashed line) are normalised to 1. The signal is shown for $M_{H} = 120$ GeV/c$^{2}$.}
\label{hbbselep1} 
\end{figure}

\begin{figure}
\begin{center}
\epsfig{figure=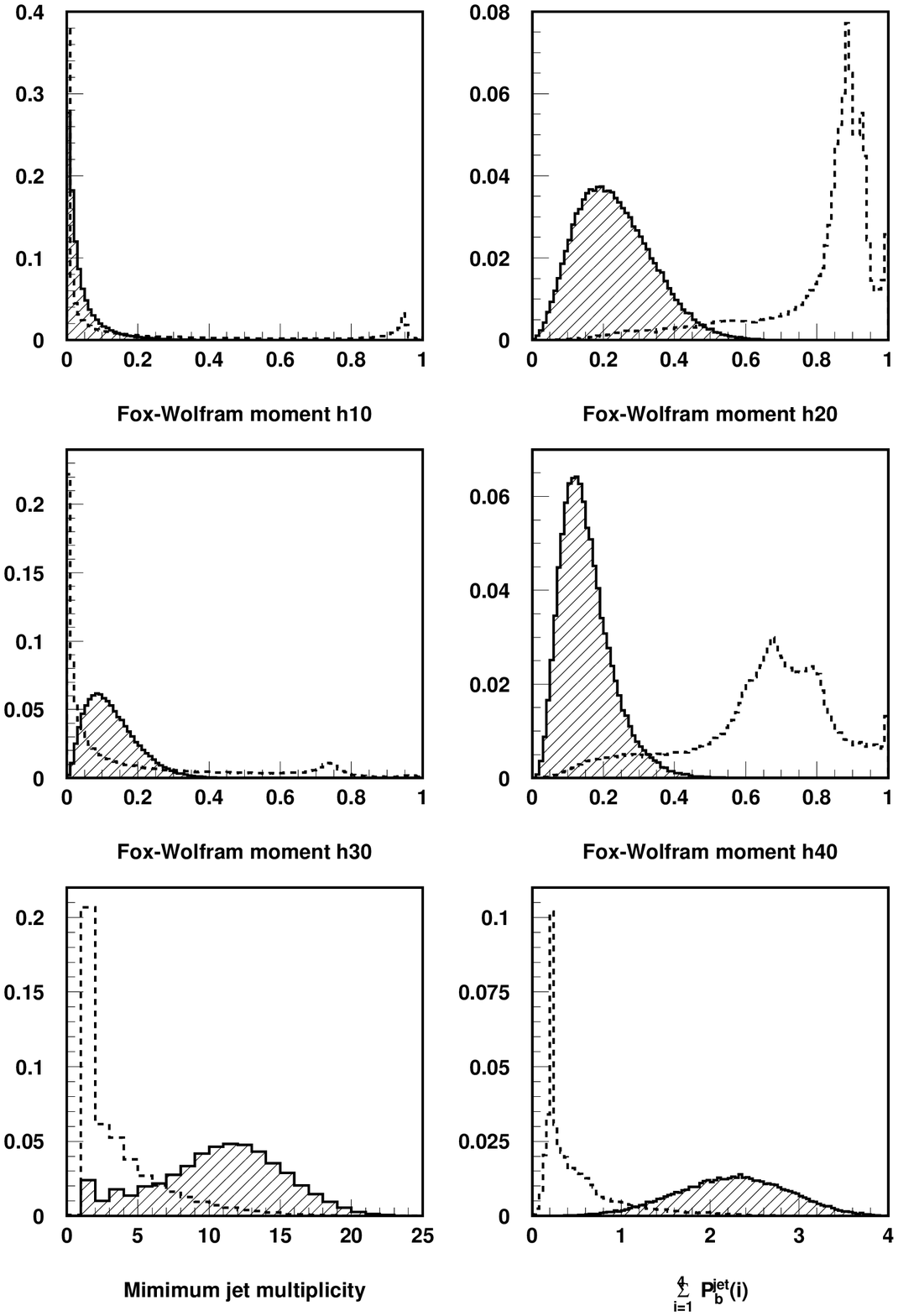,height=17cm,width=15cm}
\end{center}
\caption{\it  The $H \rightarrow b \overline b$ semileptonic channel: preselection variables (II). The signal (solid line) and the background (dashed line) are normalised to 1. The signal is shown for $M_{H} = 120$ GeV/c$^{2}$.}
\label{hbbselep2} 
\end{figure}

First, we require the presence in the event of at least one charged lepton (a $\mu^{\pm}$ or a $e^{\pm}$). Then, we request (figure~\ref{hbbselep1} and~\ref{hbbselep2}\footnote{ The figures \ref{plotresults} to \ref{effect_mtop2} are placed after the references.}):

\begin{itemize}
\item 
500 GeV/c$^{2}$$ < $ Total visible mass $ < $ 750 GeV/c$^{2}$
\item 
Total multiplicity $ \ge $ 110
\item 
Number of jets (including possible isolated leptons; JADE algorithm with $y_{cut}=1.10^{-3}$) $\ge$ 7
\item 
Thrust $\le$ 0.85
\item 
Light jet mass $\ge$ 50 GeV/c$^{2}$
\item 
Heavy jet mass $\ge$ 150 GeV/c$^{2}$
\item 
Fox-Wolfram moment h10 $\le$ 0.2 
\item 
Fox-Wolfram moment h20 $\le$ 0.6
\item 
Fox-Wolfram moment h30 $\le$ 0.4
\item 
Fox-Wolfram moment h40 $\le$  0.5

\end{itemize}

Now, an energetic and isolated charged lepton has to be identified. Among all the charged leptons ($\mu^{\pm}$ and $e^{\pm}$) reconstructed by the detector, the one which maximises : $E_{l}*(1-cos \theta)$, $E_{l}$ being the energy of the lepton and $\theta$ the angle between its direction and that of the closest jet when we force the rest of particles in a 6 jet configuration (with JADE), is chosen. This procedure allows to choose a lepton from a leptonic decay of a W rather than from a B (or D) -meson decay which tends to be less energetic and isolated.

Once the lepton is tagged, the remaining particles are forced into 6 jets with the JADE algorithm. We then flavour-tag the jets and finally we require (figure~\ref{hbbselep2}):

\begin{itemize}
\item
Mininum jet multiplicity $\ge$ 3
\item 
$\sum_{i=1}^{4} P^{jet}_{b}(i) \ge$ 1
\end{itemize}

The overall preselection efficiencies and corresponding effective cross-sections\footnote{The effective cross-section of a process X is defined as the remaining cross-section of this process after a selection procedure: $\sigma_{X}^{eff} = \sigma_{X}^{tot}*\epsilon_{X}$; where $\epsilon_{X}$ is the selection efficiency for the process X.} are shown in table~\ref{preseleffselep}. Apart from hadronic $t\overline tH$ and $t\overline tH ( H \nrightarrow b\overline b)$ events, the backgrounds with the highest preselection efficiencies are, as expected, those of the $t\overline t$ and $t\overline tZ$ productions. However, due to its relatively high cross-section, the main background after this preselection is by far the one due to the $e^{+}e^{-} \rightarrow t\overline t$ process.

\clearpage

\begin{table}[h]
\centering
\begin{tabular}{||c||c||c||c|}
\hline
\hline
{Final state} & $M_{H}$ (GeV/c$^{2})$   &{$\epsilon_{presel}$ ($\%$)}& {$\sigma _{eff} (fb)$}\\
\hline
{$t\overline tH \rightarrow 4b2ql\nu$} & {  120 } & {$61.8$} & {4.56$\cdot10^{-1}$}\\
{$t\overline tH \rightarrow 4b2ql\nu$} & {  130 } & {$62.6$} & {3.04$\cdot10^{-1}$}\\
{$t\overline tH \rightarrow 4b2ql\nu$} & {  140 } & {$64.0$} & {1.75$\cdot10^{-1}$}\\
{$t\overline tH \rightarrow 4b2ql\nu$} & {  150 } & {$65.3$} & {7.57$\cdot10^{-2}$}\\
\hline
{$ t\overline t$ } &{-} & {$9.02$} & {26.8}\\
{$ t\overline tZ$ } &{-} & {$23.5$} & {1.01}\\
{$ WW$ } &{-} & {$8.30\cdot10^{-3}$} & {3.56$\cdot10^{-1}$}\\
{$ ZZ$ } &{-} & {$7.27\cdot10^{-2}$} & {1.74$\cdot10^{-1}$}\\
{$ q\overline q$ } &{-} & {$6.25\cdot10^{-3}$} & {9.77$\cdot10^{-2}$}\\
{$t\overline tH \rightarrow 4b4q$} & { 120 } & {$40.3$} & {3.08$\cdot10^{-1}$}\\
{$t\overline tH \rightarrow 4b4q$} & { 130 } & {$40.6$} & {2.05$\cdot10^{-1}$}\\
{$t\overline tH \rightarrow 4b4q$} & { 140 } & {$40.2$} & {1.14$\cdot10^{-1}$}\\
{$t\overline tH \rightarrow 4b4q$} & { 150 } & {$40.4$} & {4.86$\cdot10^{-2}$}\\
{$t\overline tH \rightarrow 4b2l2\nu$} & { 120 } & {$7.28$} & {1.34$\cdot10^{-2}$}\\
{$t\overline tH \rightarrow 4b2l2\nu$} & { 130 } & {$8.21$} & {9.54$\cdot10^{-3}$}\\
{$t\overline tH \rightarrow 4b2l2\nu$} & { 140 } & {$9.44$} & {6.18$\cdot10^{-3}$}\\
{$t\overline tH \rightarrow 4b2l2\nu$} & { 150 } & {$10.7$} & {2.97$\cdot10^{-3}$}\\
{$t\overline tH (H \nrightarrow 2b)$} & { 120 } & {$30.3$} & {2.48$\cdot10^{-1}$}\\
{$t\overline tH (H \nrightarrow 2b)$} & { 130 } & {$31.9$} & {3.36$\cdot10^{-1}$}\\
{$t\overline tH (H \nrightarrow 2b)$} & { 140 } & {$33.3$} & {4.21$\cdot10^{-1}$}\\
{$t\overline tH (H \nrightarrow 2b)$} & { 150 } & {$33.5$} & {4.65$\cdot10^{-1}$}\\
\hline
{ Total background} &{ 120} & {$4.52\cdot10^{-1}$} & {28.9}\\
{ Total background} &{ 130} & {$4.52\cdot10^{-1}$} & {28.9}\\
{ Total background} &{ 140} & {$4.51\cdot10^{-1}$} & {28.9}\\
{ Total background } &{ 150} & {$4.51\cdot10^{-1}$} & {28.9}\\
\hline
\hline
\end{tabular}
\caption{\it The $H \rightarrow b\overline b$ semileptonic channel: preselection efficiencies ($\epsilon_{presel} $) and corresponding effective cross-sections ($\sigma _{eff}$) for various Higgs boson masses. The 4 top lines stand for the selected signal final state, while the next lines break down the different background components (including those due to $t\overline tH$ events with another Higgs decay channel).}
\label{preseleffselep}
\end{table}

\subsubsection{Neural network analysis}

The distributions used for the preselection procedure still contain unexploited information. Some of them are then used to train a neural network together with other variables related to b-tagging, to the charged lepton and including the missing momentum.

\begin{figure}
\begin{center}
\epsfig{figure=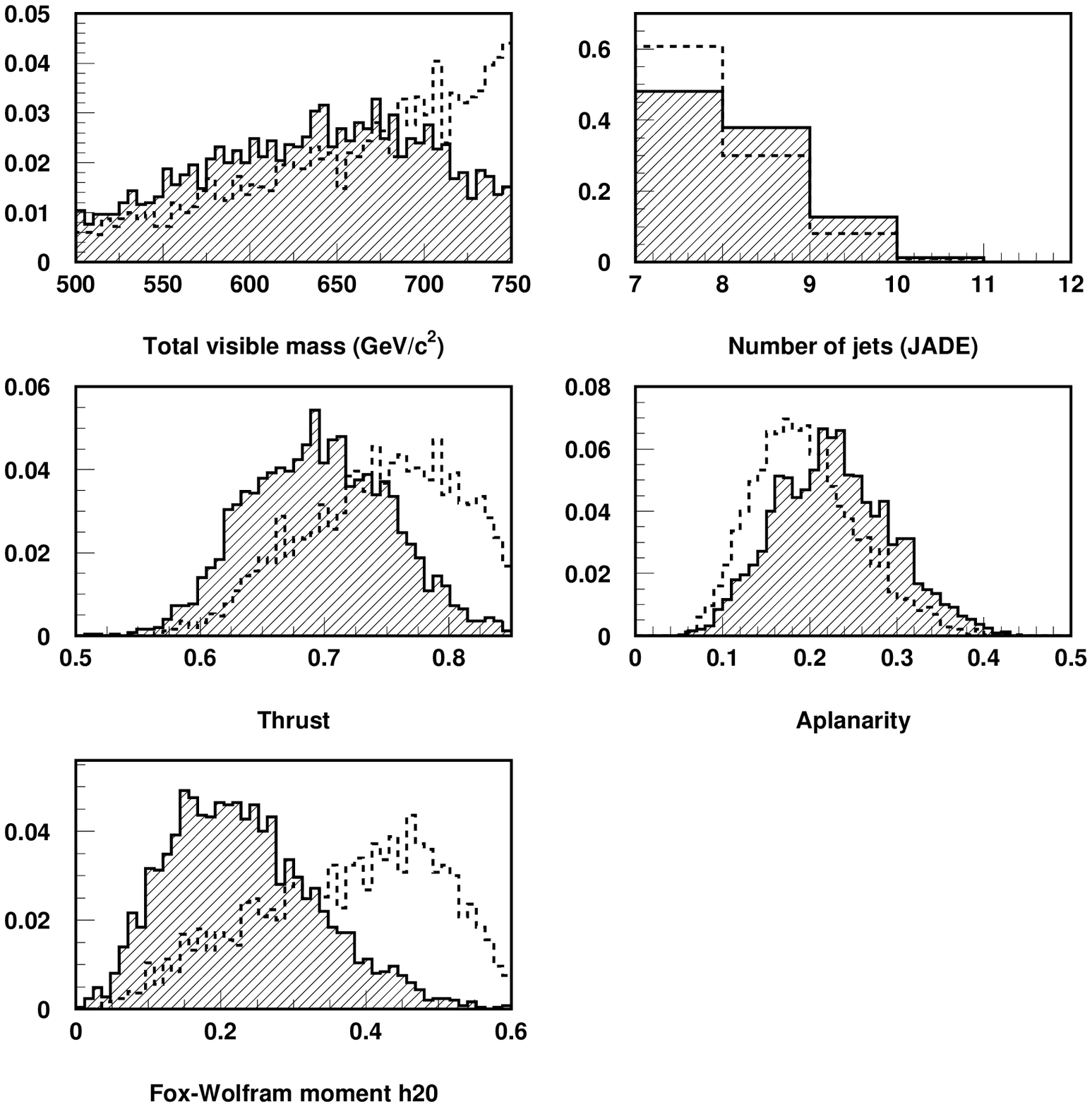,height=17cm,width=15cm}
\end{center}
\caption{\it The $H \rightarrow b \overline b$ semileptonic channel: variables (after preselection) used for the neural network analysis (I). The signal (solid line) and the background (dashed line) are normalised to 1. The signal is shown for $M_{H} = 120$ GeV/c$^{2}$.}
\label{hbbselep3} 
\end{figure}

\begin{figure}
\begin{center}
\epsfig{figure=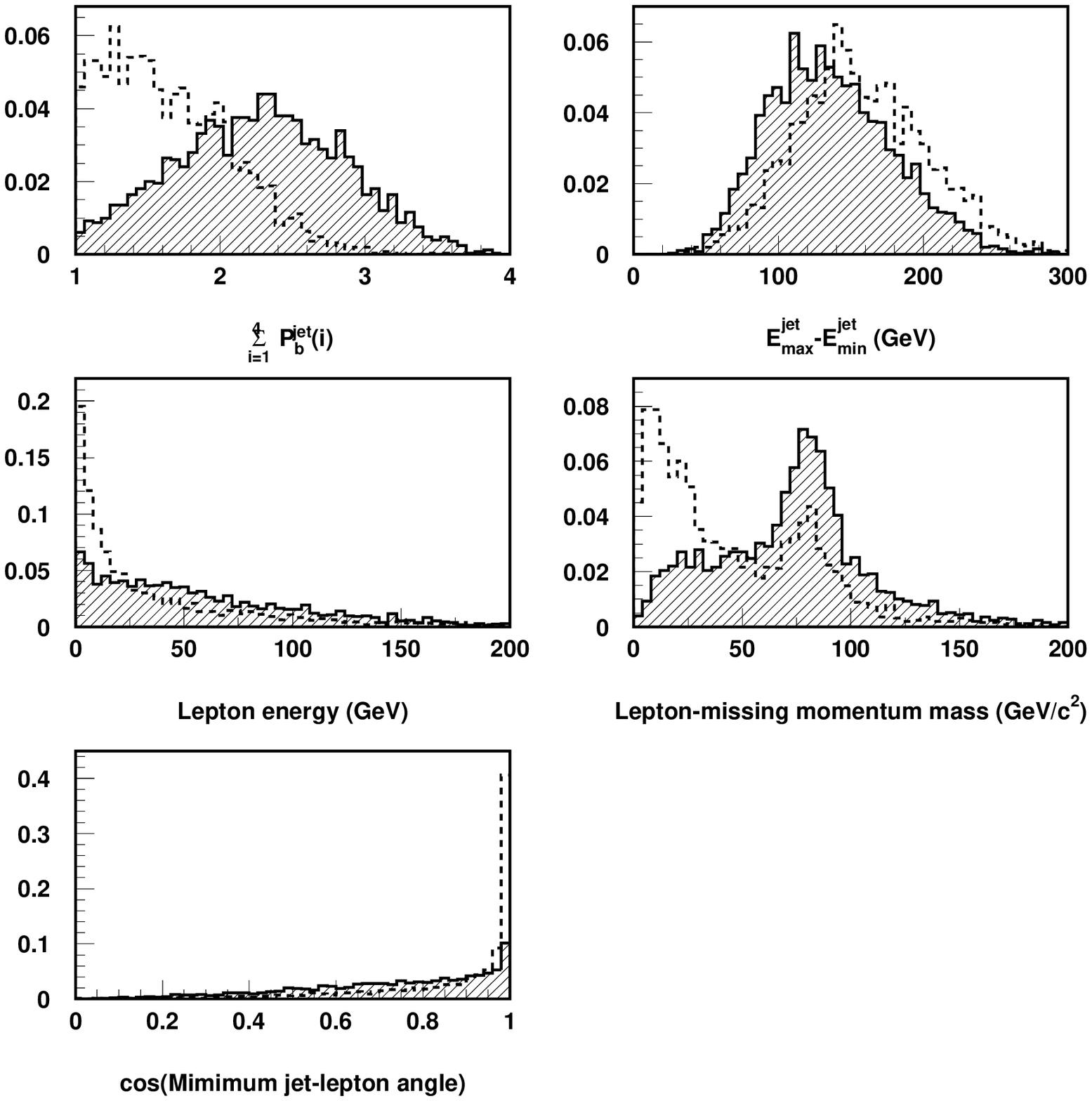,height=17cm,width=15cm}
\end{center}
\caption{\it The $H \rightarrow b \overline b$ semileptonic channel: variables (after preselection) used for the neural network analysis (II). The signal (solid line) and the background (dashed line) are normalised to 1. The signal is shown for $M_{H} = 120$ GeV/c$^{2}$.}
\label{hbbselep4} 
\end{figure}

The complete list of variables is (figure \ref{hbbselep3} and \ref{hbbselep4}):

\begin{itemize}
\item 
The total visible mass.
\item 
The number of jets (including possible isolated leptons; JADE algorithm with $y_{cut}=1.10^{-3}$).
\item 
The thrust.
\item 
The aplanarity.
\item
The second Fox-Wolfram moment h20.
\item 
$\sum_{i=1}^{4} P^{jet}_{b}(i)$.
\item 
$E^{jet}_{max}-E^{jet}_{min}$.
\item 
The energy of the tagged lepton.
\item 
The invariant mass of the system made of the lepton and the missing momentum.
\item 
The cosine of the angle between the tagged lepton and the closest jet directions.
\end{itemize}

Apart from the 5 first ones, all these variables were calculated once the event was forced to the lepton tagged plus 6 jet configuration.\\

\begin{figure}
\begin{center}
\epsfig{figure=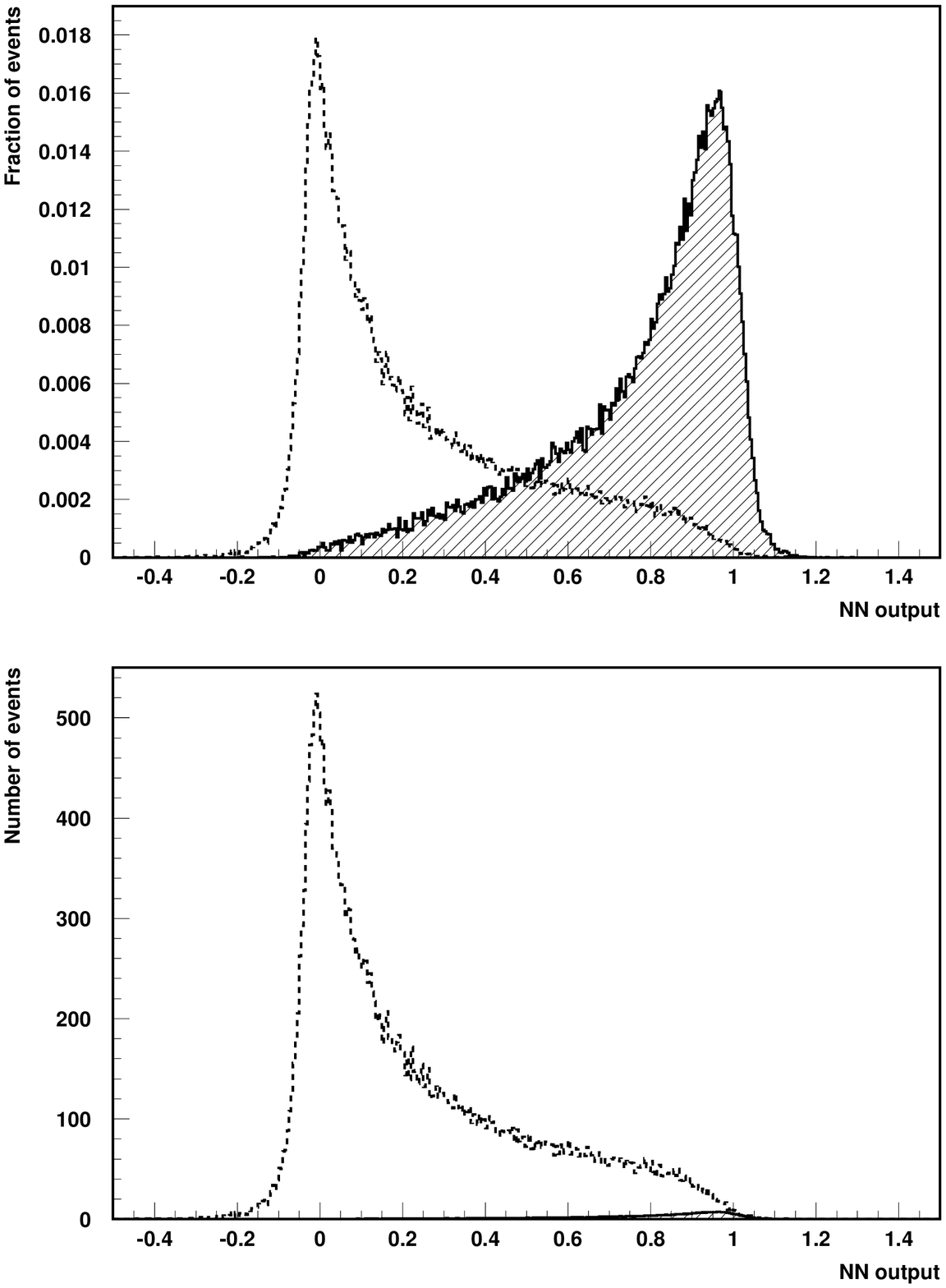,height=17cm,width=15cm}
\end{center}
\caption{\it The $H \rightarrow b \overline b$ semileptonic channel: neural network output. Top, the signal (solid line) and the background (dashed line) are normalised to 1. Bottom, the signal (solid line) and the background (dashed line) are normalised to the expected number of events. The signal is shown for $M_{H} = 120$ GeV/c$^{2}$.}
\label{hbbselep5} 
\end{figure}

Once the neural network is trained, its weights are optimized for the separation of signal from background events. The distribution of the neural network output for these 2 classes of events is shown in figure~\ref{hbbselep5} for the case where $M_{H}=120$ GeV$/c^{2}$. From this figure, we observe that the separation is quite effective. However, due to the tiny cross-section of the signal process, the purity of the selected sample and the selection efficiency of the signal will be small and, eventually, the accuracy on the measurement of $g_{ttH}$ will be limited accordingly (see relations 3.1 and 3.2).

\subsubsection{Results}

\begin{figure}[h]
\begin{center}
\psfig{figure=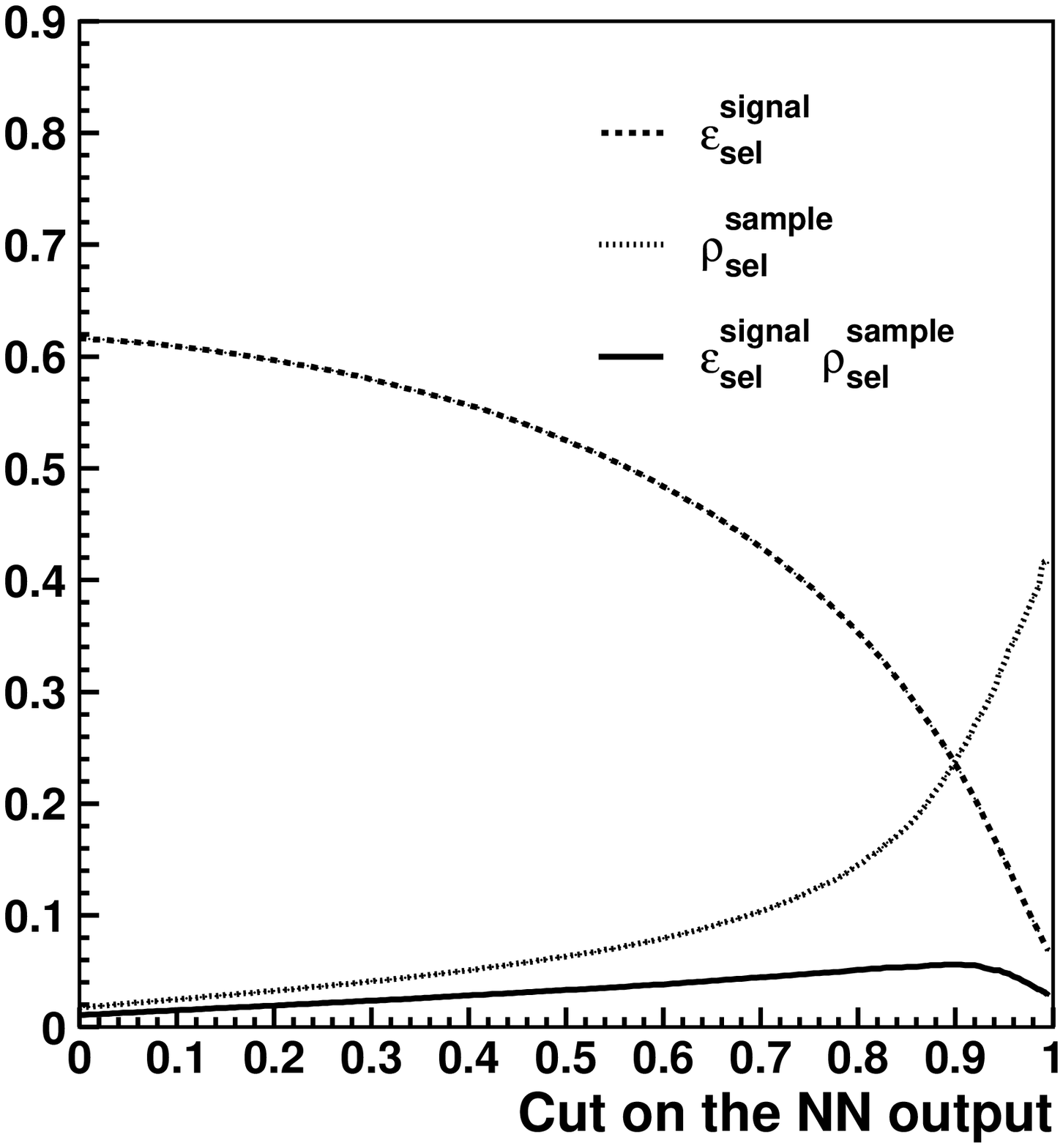,height=7cm,width=7cm}
\psfig{figure=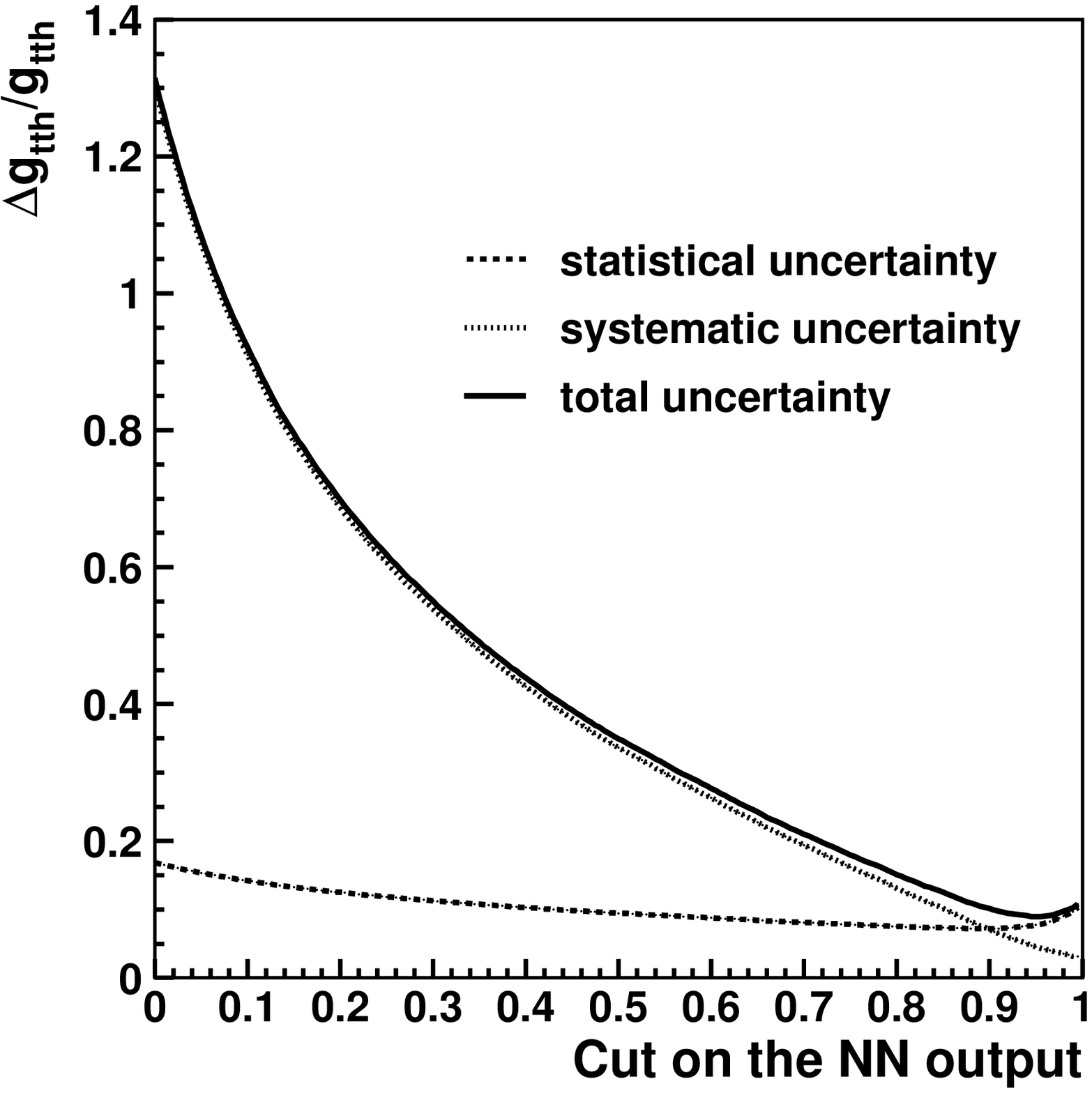,height=7cm,width=7cm}
\end{center}
\caption{\it The $H \rightarrow b\overline b$ semileptonic channel: (left) selection efficiency, purity and their product, (right) statistical, systematic and total uncertainties on the measurement of $g_{ttH}$ as functions of the value of the cut on the neural network output for $\frac {\Delta \sigma _{BG}^{eff}}{\sigma _{BG}^{eff}}$ = $5\%$  and $M_{H}=120$ GeV/c$^{2}$.}
\label{uncer_f_cut} 
\end{figure}

The next step consists to apply a cut on the output of the neural network to further separate the signal and the background. The cut value is chosen such that it minimises the quadratic sum of the statistical and systematic uncertainties. The value of this optimal cut depends on the assumption made for the value of the uncertainty on the residual background normalisation and on the number of signal events, which in turn depends on the Higgs boson mass.

The evolution of the selection efficiency of the signal $\epsilon_{sel}^{signal}$, the purity of the selected sample $\rho_{sel}^{sample}$, their products, the statistical, systematic and total uncertainties as a function of the cut on the neural network output are shown in figure~\ref{uncer_f_cut} for the case where $M_{H}=120$ GeV/c$^{2}$ and $\frac{\Delta \sigma _{BG}^{eff}}{\sigma _{BG}^{eff}} = 5\%$. As we increase the cut, the purity gets higher and its derivative is increasing. The systematic uncertainty behaves the same way as  $\frac{1-\rho_{sel}^{sample}}{\rho_{sel}^{sample}}$. It is a decreasing function of $\rho$, which varies fast for values of $\rho$ close to 0 and less as $\rho$ increases. The evolution of the selection efficiency is opposite to the one of the purity. Their product has thus a small variation. It however increases until a particular cut value (0.92 for this case). The statistical uncertainty will thus slowly decrease until this cut value. A higher cut provokes a sharp drop of the efficiency and a degradation of the statistical uncertainty. Eventually, the optimal cut is about 0.95 for the case under study.

The selection efficiencies and the corresponding effective cross-sections are shown in table \ref{seleffselep} for $M_{H}=120$ GeV/c$^{2}$ and $\frac {\Delta \sigma _{BG}^{eff}}{\sigma _{BG}^{eff}}$ = $5\%$. In this case, the number of selected events amounts to 107 for the signal and to 192 for the background. The main background after selection is due to top-pair production.

The uncertainties on $g_{ttH}$ are shown in table~\ref{selepresults} and on the figure \ref{plotresults}. They range from 9.1\% (11.7\%) to 48.7\% (65.9\%) for $\frac{\Delta \sigma _{BG}^{eff}}{\sigma _{BG}^{eff}} = 5\%$ (10\%). The resolution degrades with increasing mass due to the reduction of the statistics of the signal.

\begin{table}[h!]
\centering
\begin{tabular}{||c||c||c|}
\hline
{ Final state} &{$\epsilon_{sel} $ ($\%$)}& {$\sigma _{eff}$ $(fb)$}\\
\hline
{$t\overline tH \rightarrow 4b2ql\nu$} & {$14.6$} & {1.07$\cdot10^{-1}$}\\
\hline
{$ t\overline t$ } & {$4.26\cdot10^{-2}$} & {1.27$\cdot10^{-1}$}\\
{$ t\overline tZ$ } & {9.36$\cdot10^{-1}$} & {4.01$\cdot10^{-2}$}\\
{$ WW$ } & {$<3.7\cdot10^{-5}$} & {$<1.6\cdot10^{-3}$}\\
{$ ZZ$ } & {$5.77\cdot10^{-4}$} & {1.38$\cdot10^{-3}$}\\
{$ q\overline q$ } & {$<10^{-4}$} & {$<1.6\cdot10^{-3}$}\\
{$t\overline tH \rightarrow 4b4q$} & {$1.82$} & {1.40$\cdot10^{-2}$}\\
{$t\overline tH \rightarrow 4b2l2\nu$} & {$1.32$} & {2.33$\cdot10^{-3}$}\\
{$t\overline tH (H \nrightarrow 2b)$} & {$9.6$$\cdot10^{-1}$} & {7.86$\cdot10^{-3}$}\\
\hline
{Total background} & {$3.01\cdot10^{-3}$} & {1.92$\cdot10^{-1}$}\\
\hline
\end{tabular}
\caption{\it The $H \rightarrow b\overline b$ semileptonic channel: selection efficiencies ($\epsilon_{sel} $) and corresponding effective cross-sections ($\sigma _{eff}$) for $\frac {\Delta \sigma _{BG}^{eff}}{\sigma _{BG}^{eff}}$ = $5\%$  and $M_{H}=120$ GeV/c$^{2}$.}
\label{seleffselep}
\end{table}

\begin{table}[h!]
\centering
\begin{tabular}{||c||c||c||c||c||c||c|}
\hline
{ $M_{H}$ (GeV/c$^{2}$)} &{$\frac{\Delta \sigma _{BG}^{eff}}{\sigma _{BG}^{eff}}$} & {$\epsilon_{sel}^{signal} $}&{$\rho_{sel}^{sample}$}&{$\big(\frac{\Delta g_{ttH}}{g_{ttH}}\big)_{stat}$} & {$\big(\frac{\Delta g_{ttH}}{g_{ttH}}\big)_{syst}$} &  {$\frac{\Delta g_{ttH}}{g_{ttH}}$}\\
\hline
\hline
{ 120} &{$5\%$} &{$14.6\%$}&{$36.0\%$}&{$8.0\%$}&{$4.5\%$}&{$9.1\%$}\\
{    } &{$10\%$}&{$10.7\%$}&{$39.7\%$} &{$8.8\%$}&{$7.7\%$}&{$11.7\%$}\\
\hline
{ 130} &{$5\%$} &{$16.3\%$}&{$26.0\%$}&{$10.9 \%$}&{$7.2\%$}&{$13.0\%$}\\
{    } &{$10\%$}&{$12.9\%$}&{$28.8\%$} &{$11.6\%$}&{$12.5\%$}&{$17.1\%$}\\
\hline
{ 140} &{$5\%$} &{$17.4\%$}&{$17.1\%$}&{$17.5\%$}&{$12.2\%$}&{$21.3\%$}\\
{    } &{$10\%$}&{$11.1\%$}&{$20.1\%$ } &{$20.2\%$}&{$20.1\%$}&{$28.5\%$}\\
\hline
{ 150} &{$5\%$} &{$13.8\%$}&{$9.0\%$}&{$41.4\%$}&{$25.6\%$}&{$48.7\%$}\\
{    } &{$10\%$}&{$13.8\%$}&{$9.0\%$} &{$41.4\%$}&{$51.2\%$}&{$65.9\%$}\\
\hline
\hline
\end{tabular}
\caption{\it The $H \rightarrow b\overline b$ semileptonic channel: expected uncertainty on the measurement of $g_{ttH}$ for various Higgs boson masses. Selection efficiency of the signal ($\epsilon_{sel}^{signal} $) and purity of the selected sample ($\rho_{sel}^{sample}$) are also shown.}
\label{selepresults}
\end{table}

\subsubsection{Precision loss due to 6 fermion processes}

In the case of the final state under study, the most contaminating 6 fermion processes are the ones  made of at least 4 b-quarks. They were studied for $M_{H}=120$ GeV/c$^{2}$. Their cross section and selection efficiency are shown in table~\ref{seleffselep6f} and the loss of resolution on the measurement of $g_{ttH}$ is shown in table~\ref{selepresults6f}. One observes that it can be neglected.

\begin{table}[h]
\centering
\begin{tabular}{||c|c|c||}
\hline
{ Final state} & {$\sigma  (fb)$}& {$\epsilon_{sel} $}\\
\hline
\hline
{$b\overline bb\overline bb\overline b$ } & {$6.4\cdot10^{-3}$} & {$5.0\%$}\\
{$b\overline bb\overline bq\overline q$ $(q=u,d)$ } & {$1.2\cdot10^{-1}$}& {$1.0\%$}\\
{$b\overline bb\overline b s\overline s$ } & {$6.3\cdot10^{-2}$} & {$1.3\%$}\\
{$b\overline bb\overline b c\overline c$ } & {$5.1\cdot10^{-2}$} & {$2.0\%$}\\
\hline
{$b\overline bb\overline b t\overline t$ } & {$6.9\cdot10^{-3}$}  & {$14.2\%$}\\
{$b\overline bq\overline q t\overline t $ $(q=u,d,s)$} & {$8.9\cdot10^{-3}$} & {$5.2\%$} \\
{$b\overline bc\overline c t\overline t $ } & {$3.5\cdot10^{-3}$} & {$6.6\%$}\\
{$b\overline bt\overline b d\overline u$ $^{*}$} & {$\sim 1\cdot10^{-2}$} & {$5.4\%$}\\
{$b\overline bt\overline b \mu \overline \nu$ $^{*}$} & {$\sim 3\cdot10^{-3}$}  & {$14.2\%$}\\
\hline
\end{tabular}
\caption{\it The $H \rightarrow b\overline b$ semileptonic channel: cross section ($\sigma$) and selection efficiencies ($\epsilon_{sel} $) of the 6 fermion processes for $\frac{\Delta \sigma _{BG}^{eff}}{\sigma _{BG}^{eff}}$ = $5\%$ and $M_{H}=120$ GeV/c$^{2}$. For events marked with a $^{*}$, some diagrams were removed from the calculation to avoid double counting.}
\label{seleffselep6f}
\end{table}

\begin{table}[h]
\centering
\begin{tabular}{||c||c||c|}
\hline
{$\frac{\Delta \sigma _{BG}^{eff}}{\sigma _{BG}^{eff}}$}  &  {$\frac{\Delta g_{ttH}}{g_{ttH}}$}(without 6f) &  {$\frac{\Delta g_{ttH}}{g_{ttH}}$(with 6f)}\\
\hline
\hline
{$5\%$} &{$  9.1\%$} & {$9.3\%$}\\
{$10\%$} &{$ 11.7\%$} & {$12.1\%$}\\
\hline
\end{tabular}
\caption{\it The $H \rightarrow b\overline b$ semileptonic channel: expected uncertainty on the measurement of $g_{ttH}$ without and with inclusion of the 6 fermion processes in the analysis for $M_{H}=120$ GeV/c$^{2}$.}
\label{selepresults6f}
\end{table}

\subsection{Hadronic channel}

\subsubsection{Introduction}

The final state follows from the process:
\\

$e^{+}e^{-} \rightarrow t\overline tH \rightarrow W^{+}bW^{-}\overline b b\overline b \rightarrow 4b4q$.
\\

This channel is thus characterized by 4 b-jets and 4 light quark jets. It benefits from the highest branching ratio. However the environnement is still more crowded than in the semileptonic channel and the event reconstruction will be quite affected.  Notice that, here, the semileptonic channel ($t\overline t H \rightarrow 4b2ql\nu$), the fully leptonic channel ($t\overline t H \rightarrow 4b2l2\nu$) and the channels where the Higgs boson doesn't decay into b quarks act as background processes.

\subsubsection{Sequential analysis}

We first require:

\begin{itemize}

\item 
Total visible mass $\ge$ 560 GeV/c$^{2}$.
\item 
Total multiplicity $\ge$ 120.
\item 
Number of jets (including possible isolated leptons; JADE algorithm with $y_{cut}=1.10^{-3}$) $\ge$ 7.
\item 
Thrust $\le$ 0.85.
\end{itemize}

The events satisfying these requirements are forced in a 8 jet configuration (using JADE) and a second set of criteria is then applied:\\
\begin{itemize}

\item 

Minimum jet multiplicity $>$ 1.
\item 
Minimum dijet invariant mass $\ge$ 15 GeV/c$^{2}$.
\item 
$E^{jet}_{min}$ $\ge$ 20 GeV.

\end{itemize}

\begin{figure}
\begin{center}
\psfig{figure=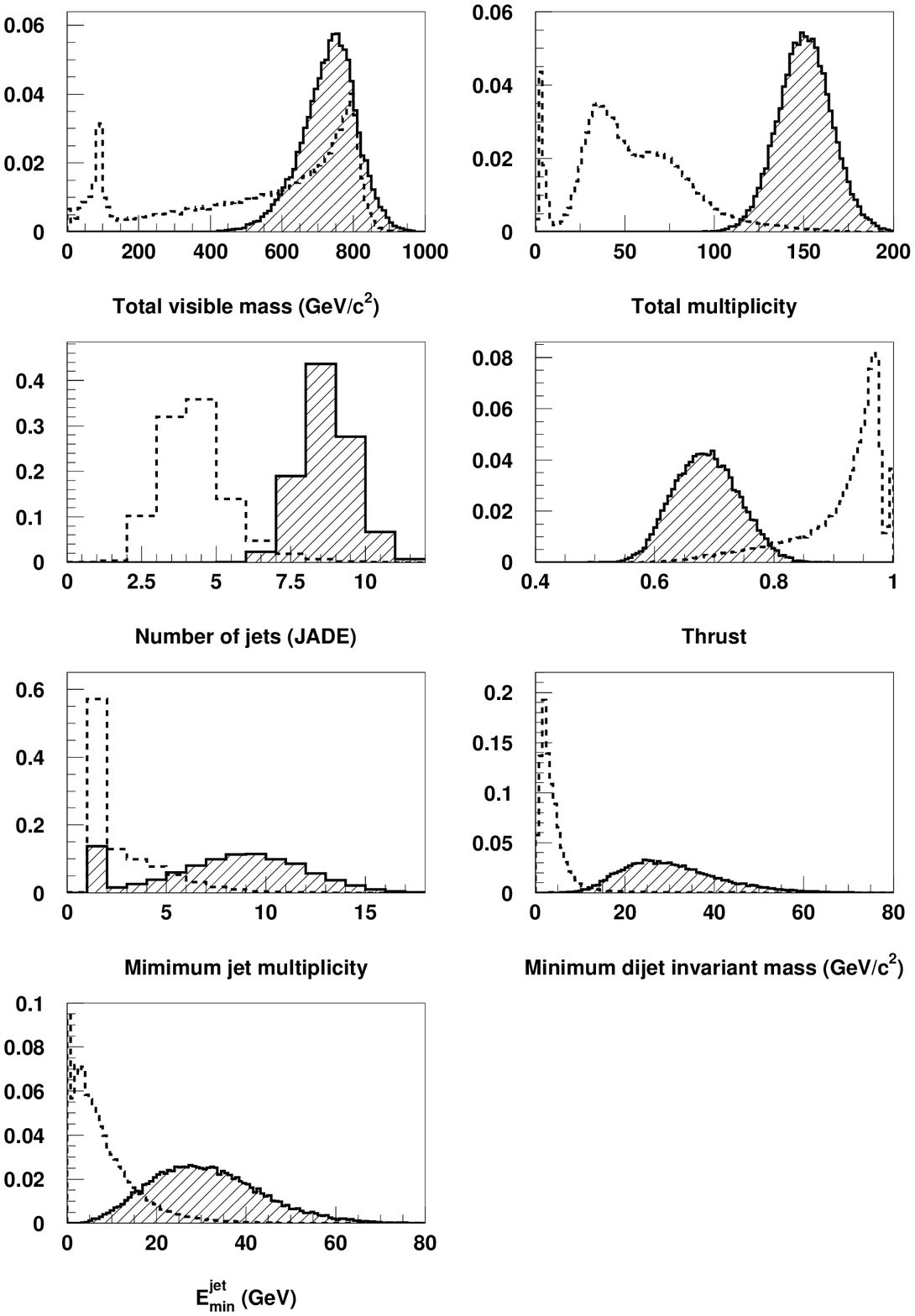,height=18cm,width=15cm}
\end{center}
\caption{\it The $H \rightarrow b \overline b$ hadronic channel: preselection variables. The signal (solid line) and the background (dashed line) are normalised to 1. The signal is shown for $M_{H} = 120$ GeV/c$^{2}$.}
\label{hbbhad1} 
\end{figure}

All these distributions are shown in figure~\ref{hbbhad1}. The overall preselection efficiencies and the corresponding effective cross-sections are shown in table~\ref{preseleffhadro}. The remarks made on the residual background in the semileptonic channel (see section 5.2.2) apply here as well.

\begin{table}[h!]
\centering
\begin{tabular}{||c||c||c||c|}
\hline
\hline
{ Final state} & {$M_{H}$ GeV/c$^{2}$} & {$\epsilon_{presel} $ ($\%$)} & {$\sigma _{eff} (fb)$}\\
\hline
{$t\overline tH \rightarrow 4b4q$} & {120} & {$63.2$} & {4.85$\cdot10^{-1}$}\\
{$t\overline tH \rightarrow 4b4q$} & {130} & {$64.0$} & {3.26$\cdot10^{-1}$}\\
{$t\overline tH \rightarrow 4b4q$} & {140} & {$64.3$} & {1.81$\cdot10^{-1}$}\\
{$t\overline tH \rightarrow 4b4q$} & {150} & {$64.6$} & {7.84$\cdot10^{-2}$}\\
\hline
{$ t\overline t$ } & {-} & {$7.37$} & {21.9}\\
{$ t\overline tZ$ } & {-} & {$22.8$} & {9.74$\cdot10^{-1}$}\\
{$ WW$ } & {-} & {1.10$\cdot10^{-1}$} & {4.74}\\
{$ ZZ$ } & {-} & {2.56$\cdot10^{-1}$} & {6.14$\cdot10^{-1}$}\\
{$ q\overline q$ } & {-} & {$6.93\cdot10^{-2}$} & {1.08}\\
{$t\overline tH \rightarrow 4b2ql\nu $ } & {120} & {$14.1$} & {1.04$\cdot10^{-1}$}\\
{$t\overline tH \rightarrow 4b2ql\nu $ } & {130} & {$14.7$} & {7.20$\cdot10^{-2}$}\\
{$t\overline tH \rightarrow 4b2ql\nu $ } & {140} & {$16.0$} & {4.32$\cdot10^{-2}$}\\
{$t\overline tH \rightarrow 4b2ql\nu $ } & {150} & {$16.4$} & {1.91$\cdot10^{-2}$}\\
{$t\overline tH \rightarrow 4b2l2\nu$ } & {120} & {$3.60$$\cdot10^{-1}$} & {6.66$\cdot10^{-4}$}\\
{$t\overline tH \rightarrow 4b2l2\nu$ } & {130} & {$3.60$$\cdot10^{-1}$} & {4.42$\cdot10^{-4}$}\\
{$t\overline tH \rightarrow 4b2l2\nu$ } & {140} & {$5.20$$\cdot10^{-1}$} & {3.52$\cdot10^{-4}$}\\
{$t\overline tH \rightarrow 4b2l2\nu$ } & {150} & {$6.85$$\cdot10^{-1}$} & {2.00$\cdot10^{-4}$}\\
{$t\overline tH (H \nrightarrow 2b)$} & { 120 } & {$26.4$} & {2.16$\cdot10^{-1}$}\\
{$t\overline tH (H \nrightarrow 2b)$} & { 130 } & {$29.5$} & {3.14$\cdot10^{-1}$}\\
{$t\overline tH (H \nrightarrow 2b)$} & { 140 } & {$30.6$} & {3.85$\cdot10^{-1}$}\\
{$t\overline tH (H \nrightarrow 2b)$} & { 150 } & {$31.9$} & {4.39$\cdot10^{-1}$}\\
\hline
{Total background } &{ 120} & {$4.46\cdot10^{-1}$} & {28.6}\\
{Total background } &{ 130} & {$4.47\cdot10^{-1}$} & {28.6}\\
{Total background } &{ 140} & {$4.48\cdot10^{-1}$} & {28.7}\\
{Total background } &{ 150} & {$4.49\cdot10^{-1}$} & {28.7}\\
\hline
\hline
\end{tabular}
\caption{\it The $H \rightarrow b\overline b$ hadronic channel: preselection efficiencies ($\epsilon_{presel} $) and corresponding effective cross-sections ($\sigma _{eff}$) for various Higgs boson masses. The 4 top lines stand for the selected signal final state, while the next lines break down the different background components (including those due to $t\overline tH$ events with another Higgs decay channel)}
\label{preseleffhadro}
\end{table}

\subsubsection{Neural network analysis}

We then train a neural network to use the information contained in the distributions in a more powerful way. The variables used are:

\begin{itemize}

\item 
The total visible mass.
\item 
The number of jets (including possible isolated leptons; JADE algorithm with $y_{cut}=1.10^{-3}$).
\item 
The thrust.
\item 
The aplanarity.
\item 
The light jet mass.
\item 
The heavy jet mass.
\item 
The second Fox-Wolfram moment h20.
\item 
$\sum_{i=1}^{4} P^{jet}_{b}(i)$.
\item 
$E^{jet}_{max}-E^{jet}_{min}$.

\end{itemize}

\begin{figure}
\begin{center}
\psfig{figure=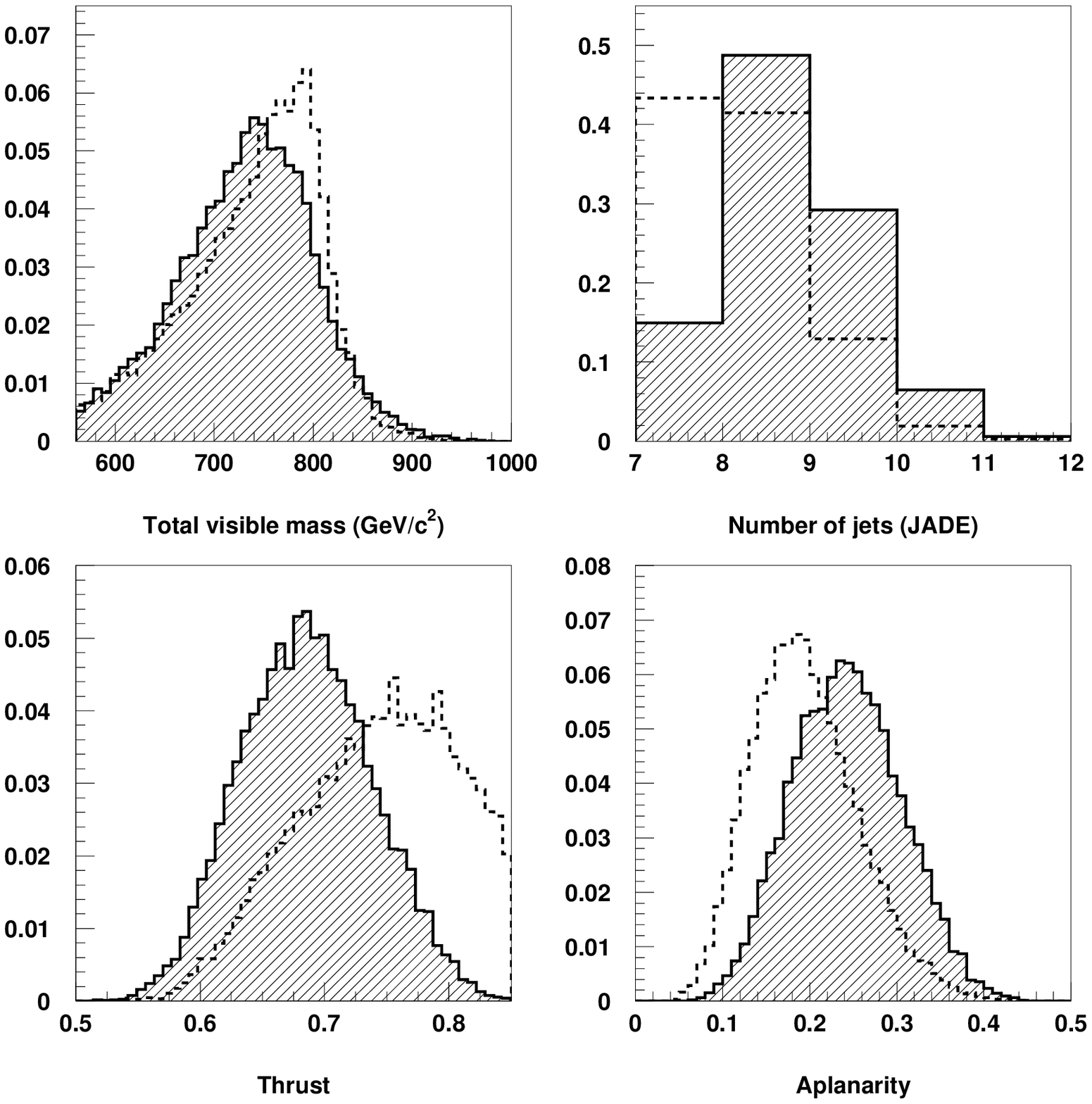,height=17cm,width=15cm}
\end{center}
\caption{\it The $H \rightarrow b \overline b$ hadronic channel: variables (after preselection) used for the neural network analysis (I). The signal (solid line) and the background (dashed line) are normalised to 1. The signal is shown for $M_{H} = 120$ GeV/c$^{2}$.}
\label{hbbhad2} 
\end{figure}

\begin{figure}
\begin{center}
\psfig{figure=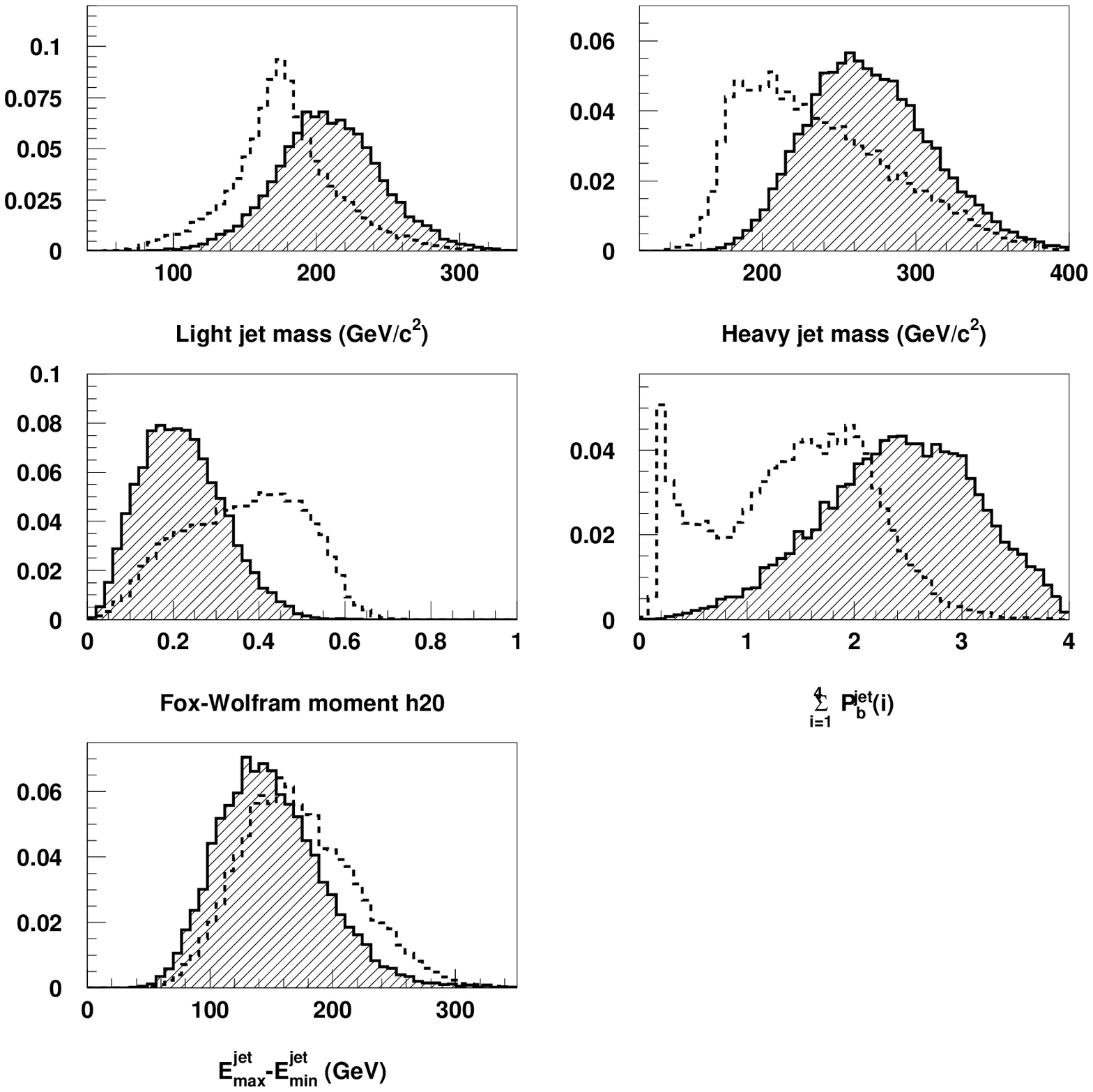,height=17cm,width=15cm}
\end{center}
\caption{\it The $H \rightarrow b \overline b$ hadronic channel: variables (after preselection) used for the neural network analysis (II). The signal (solid line) and the background (dashed line) are normalised to 1. The signal is shown for $M_{H} = 120$ GeV/c$^{2}$.}
\label{hbbhad3} 
\end{figure}

$E^{jet}_{max(min)}$ and the $P^{jet}_{b}(i)$ are calculated once the event is forced into 8 jets. The distributions of these variables (after the preselection) are shown in figures~\ref{hbbhad2} and~\ref{hbbhad3}.

\subsubsection{Results}

\begin{figure}
\begin{center}
\psfig{figure=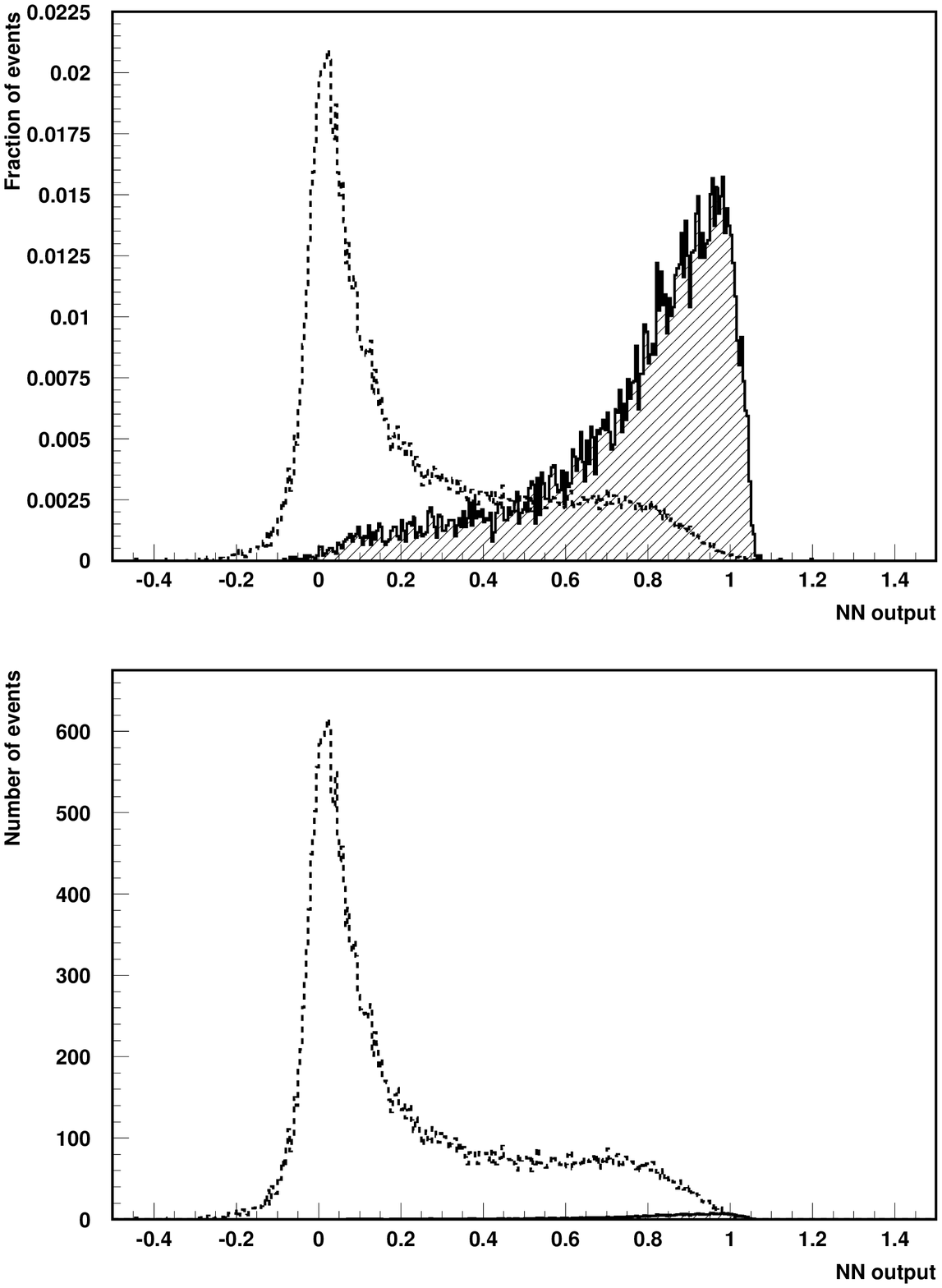,height=17cm,width=15cm}
\end{center}
\caption{\it The $H \rightarrow b \overline b$ hadronic channel: neural network output. Top, the signal (solid line) and the background (dashed line) are normalised to 1. Bottom, the signal (solid line) and the background (dashed line) are normalised to the expected number of events. The signal is shown for $M_{H} = 120$ GeV/c$^{2}$.}
\label{hbbhad5} 
\end{figure}

The result of the training is shown in figure~\ref{hbbhad5}. We apply the same procedure as for the semileptonic channel (see section 5.2.4). An example of the values of the selection efficiencies and of the corresponding effective cross-sections is shown in table~\ref{seleffhadro}. As for the semileptonic channel, the main background is due to top pair production. The resolutions on $g_{ttH}$ are shown in table~\ref{hadroresults} and on figure \ref{plotresults}. They range from 8.3\% (10.1\%) to 42.7\% (51.2\%) for $\frac{\Delta \sigma _{BG}^{eff}}{\sigma _{BG}^{eff}} = 5\%$ (10\%). As in the case of the semileptonic channel, the resolution degrades with increasing mass due to the reduction of the statistics of the signal.

\begin{table}[h]
\centering
\begin{tabular}{||c||c||c|}
\hline
\hline
{ Final state} & {$\epsilon_{sel} $ ($\%$)}& {$\sigma _{eff} (fb)$}\\
\hline
{$t\overline tH \rightarrow 4b4q$} & {$14.8$} & {1.14$\cdot10^{-1}$}\\
\hline
{$ t\overline t$ } & {$3.10\cdot10^{-2}$} & {9.29$\cdot10^{-2}$}\\
{$ t\overline tZ$ } & {$1.11$} & {4.74$\cdot10^{-2}$}\\
{$ WW$ } & {$<3.7\cdot10^{-5}$} & {$<1.6\cdot10^{-3}$}\\
{$ ZZ$ } & {$8.71\cdot10^{-4}$} & {2.08$\cdot10^{-3}$}\\
{$ q\overline q$ } & {$<10^{-4}$} & {$<1.6\cdot10^{-3}$}\\
{$t\overline tH \rightarrow  4b2ql\nu $} & {$2.44$} & {1.80$\cdot10^{-2}$}\\
{$t\overline tH \rightarrow 4b2l2\nu$} & {$3.14\cdot10^{-2}$} & {5.55$\cdot10^{-5}$}\\
{$t\overline tH (H \nrightarrow 2b)$} & {7.9$\cdot10^{-1}$} & {6.47$\cdot10^{-3}$}\\
\hline
{Total background } & {$2.61\cdot10^{-3}$} & {1.67$\cdot10^{-1}$}\\
\hline
\end{tabular}
\caption{\it The $H \rightarrow b\overline b$ hadronic channel: selection efficiencies ($\epsilon_{sel} $) and corresponding effective cross-sections ($\sigma _{eff}$) for $\frac{\Delta \sigma _{BG}^{eff}}{\sigma _{BG}^{eff}}$ = $5\%$  and $M_{H}=120$ GeV/c$^{2}$.}
\label{seleffhadro}
\end{table}

\begin{table}[h]
\centering
\begin{tabular}{||c||c||c||c||c||c||c|}
\hline
\hline
{ $M_{H}$ (GeV/c$^{2}$)} &{$\frac{\Delta \sigma _{BG}^{eff}}{\sigma _{BG}^{eff}}$} & {$\epsilon_{sel}^{signal} $}&{$\rho_{sel}^{sample}$}& {$\big(\frac{\Delta g_{ttH}}{g_{ttH}}\big)_{stat}$} & {$\big(\frac{\Delta g_{ttH}}{g_{ttH}}\big)_{syst}$} &  {$\frac{\Delta g_{ttH}}{g_{ttH}}$}\\
\hline
\hline
{ 120} &{$5\%$} &{$14.8\%$}&{$40.5\%$} &{$7.4\%$}&{$3.7\%$}&{$8.3\%$}\\
{    } &{$10\%$}&{$12.0\%$}&{$44.7\%$} &{$7.9\%$}&{$6.3\%$}&{$10.1\%$}\\
\hline
{ 130} &{$5\%$} &{$15.6\%$}&{$27.5\%$} &{$10.8\%$}&{$6.7\%$}&{$12.7\%$}\\
{    } &{$10\%$}&{$9.5\%$}&{$35.5\%$} &{$12.2\%$}&{$9.2\%$}&{$15.3\%$}\\
\hline
{ 140} &{$5\%$} &{$13.7\%$}&{$19.9\%$} &{$18.3\%$}&{$10.2\%$}&{$20.9\%$}\\
{    } &{$10\%$}&{$8.9\%$}&{$23.7\%$} &{$20.7\%$}&{$16.3\%$}&{$26.4\%$}\\
\hline
{ 150} &{$5\%$} &{$12.4\%$}&{$11.7\%$} &{$38.2\%$}&{$19.1\%$}&{$42.7\%$}\\
{    } &{$10\%$}&{$8.8\%$}&{$14.2\%$} &{$41.1\%$}&{$30.5\%$}&{$51.2\%$}\\
\hline
\hline
\end{tabular}
\caption{\it The $H \rightarrow b\overline b$ hadronic channel: expected uncertainty on the measurement of $g_{ttH}$ for various Higgs boson masses. Selection efficiency of the signal ($\epsilon_{sel}^{signal} $) and purity of the selected sample ($\rho_{sel}^{sample}$) are also shown.}
\label{hadroresults}
\end{table}

\subsubsection{Precision loss due to 6 fermion processes}

The 6 fermion processes which may degrade the resolution on $g_{ttH}$ in this channel are the same as in the $H \rightarrow b\overline b$ semileptonic channel and they were also studied for $M_{H}=120$ GeV/c$^{2}$. Their cross sections and selection efficiencies are shown in table \ref{seleffhad6f}. The precisions obtained on $g_{ttH}$ when the 6 fermion background is included are shown in table~\ref{hadroresults6f}. As in the semileptonic channel, the degradation of the measurement due to the 6 fermion background is negligible.

\begin{table}[h]
\centering
\begin{tabular}{||c|c|c||}
\hline
{Final state} & {$\sigma  (fb)$}& {$\epsilon_{sel} $}\\
\hline
\hline
{$b\overline bb\overline bb\overline b$ } & {$6.4\cdot10^{-3}$}& {$5.8\%$}\\
{$b\overline bb\overline bq\overline q$ $(q=u,d)$ } & {$1.2\cdot10^{-1}$} & {$1.6\%$}\\
{$b\overline bb\overline b s\overline s$ } & {$6.3\cdot10^{-2}$} & {$1.8\%$}\\
{$b\overline bb\overline b c\overline c$ } & {$5.1\cdot10^{-2}$} & {$2.0\%$}\\
\hline
{$b\overline bb\overline b t\overline t$ } & {$6.9\cdot10^{-3}$}  & {$31.8\%$}\\
{$b\overline bq\overline q t\overline t $ $(q=u,d,s)$} & {$8.9\cdot10^{-3}$} & {$12.5\%$} \\
{$b\overline bc\overline c t\overline t $ } & {$3.5\cdot10^{-3}$} & {$17.2\%$}\\
{$b\overline bt\overline b d\overline u$ $^{*}$} & {$\sim 1\cdot10^{-2}$}& {$14.0\%$}\\
{$b\overline bt\overline b \mu \overline \nu$  $^{*}$} & {$\sim 3\cdot10^{-3}$}  & {$1.2\%$}\\
\hline
\end{tabular}
\caption{\it The $H \rightarrow b\overline b$ hadronic channel: cross section ($\sigma$) and selection efficiencies ($\epsilon_{sel} $) of the 6 fermion processes for $\frac{\Delta \sigma _{BG}^{eff}}{\sigma _{BG}^{eff}}$ = $5\%$  and $M_{H}=120$ GeV/c$^{2}$. For events marked with a $^{*}$, some diagrams were removed from the calculation to avoid double counting.}
\label{seleffhad6f}
\end{table}

\begin{table}[h]
\centering
\begin{tabular}{||c||c||c|}
\hline
{$\frac{\Delta \sigma _{BG}^{eff}}{\sigma _{BG}^{eff}}$}  &  {$\frac{\Delta g_{ttH}}{g_{ttH}}$}(without 6f) &  {$\frac{\Delta g_{ttH}}{g_{ttH}}$(with 6f)}\\
\hline
\hline
{$5\%$} &{$8.3 \%$} & {$8.5\%$}\\
{$10\%$} &{$10.1\%$} & {$10.5\%$}\\
\hline
\end{tabular}
\caption{\it The $H \rightarrow b\overline b$ hadronic channel: expected uncertainty on the measurement of $g_{ttH}$ without and with inclusion of the 6 fermion processes in the analysis for $M_{H}=120$ GeV/c$^{2}$.}
\label{hadroresults6f}
\end{table}

\section{Study of the $H \rightarrow W^{+}W^{-}$ decay mode}

\subsection{Introduction}

As we can see from the previous section, the measurement of $g_{ttH}$ in the $H \rightarrow b\overline b$ decay mode degrades quite rapidly as the Higgs boson mass increases. The reasons are the drop of the cross-section and even more, the decrease of the branching ratio of the Higgs boson into $b\overline b$ for the benefit of the $H \rightarrow W^{+}W^{-}$ decay. We will thus try to exploit this latter mode when the Higgs boson is heavier than 140 GeV/c$^{2}$, and see whether it also improves the accuracy on $g_{ttH}$ for lighter Higgs boson masses.

In this decay mode, four intermediate W bosons are present, leading to several classes of final states. Unlike the final states where the Higgs boson decays into a pair of b-quarks, there are only two b-jets in the event, thus the b-tagging is no longer an essential point of the analysis. Therefore, particular final states have to be identified which can allow good signal and background separation and which have enough statistics. Two such channels were found: the ``2 like sign lepton plus 6 jet channel'', when two W bosons of the same sign decay leptonically whereas the two remaining ones decay hadronically, and the ``single lepton plus 8 jet channel'' when only one of the W's decays leptonically. As for channels where the Higgs boson decays into pairs of b-quarks, these final states have large particle and jet multiplicities, an isotropic topology and a tiny event rate. For each channel, the same analysis will be repeated for 4 values of the Higgs boson mass within the range 130 GeV/c$^{2}$ - 200 GeV/c$^{2}$.

\subsection{The 2 like sign lepton plus 6 jet channel}

\subsubsection{Introduction}

The final state follows from the process:
\\
$e^{+}e^{-} \rightarrow t\overline tH \rightarrow W^{+}bW^{-}\overline b W^{+}W^{-} \rightarrow 2l^{\pm}2\nu2b4q$
\\

Its branching ratio is:

\begin{eqnarray}
BR(t\overline tH \rightarrow 2l^{\pm}2\nu2b4q) & = & BR(H \rightarrow W^{+}W^{-}) \nonumber\\
& & *BR(W^{\pm}W^{\pm} \rightarrow 4q)*BR(W^{\pm}W^{\pm} \rightarrow 2l2\nu)*2 \nonumber\\
& \approx & 9.6\%*BR(H \rightarrow W^{+}W^{-})\nonumber
\end{eqnarray}

This channel is thus characterized by a missing 4-momentum, two energetic and isolated charged leptons of the same sign, 4 light quark jets and 2 b-jets. In comparison with the case where no restriction on the charged lepton signs is made, requiring the two leptons to have the same sign divides the branching ratio of the signal by a factor 3 but the background can be very effectively suppressed.

Here, $t\overline tH (H \rightarrow WW) \nrightarrow 2l^{\pm}2\nu2b4q$ and $ t\overline tH (H \nrightarrow WW)$ act as background processes.
The analysis will be purely sequential, no neural network being applied.

\subsubsection{Analysis}

\begin{figure}
\begin{center}
\epsfig{figure=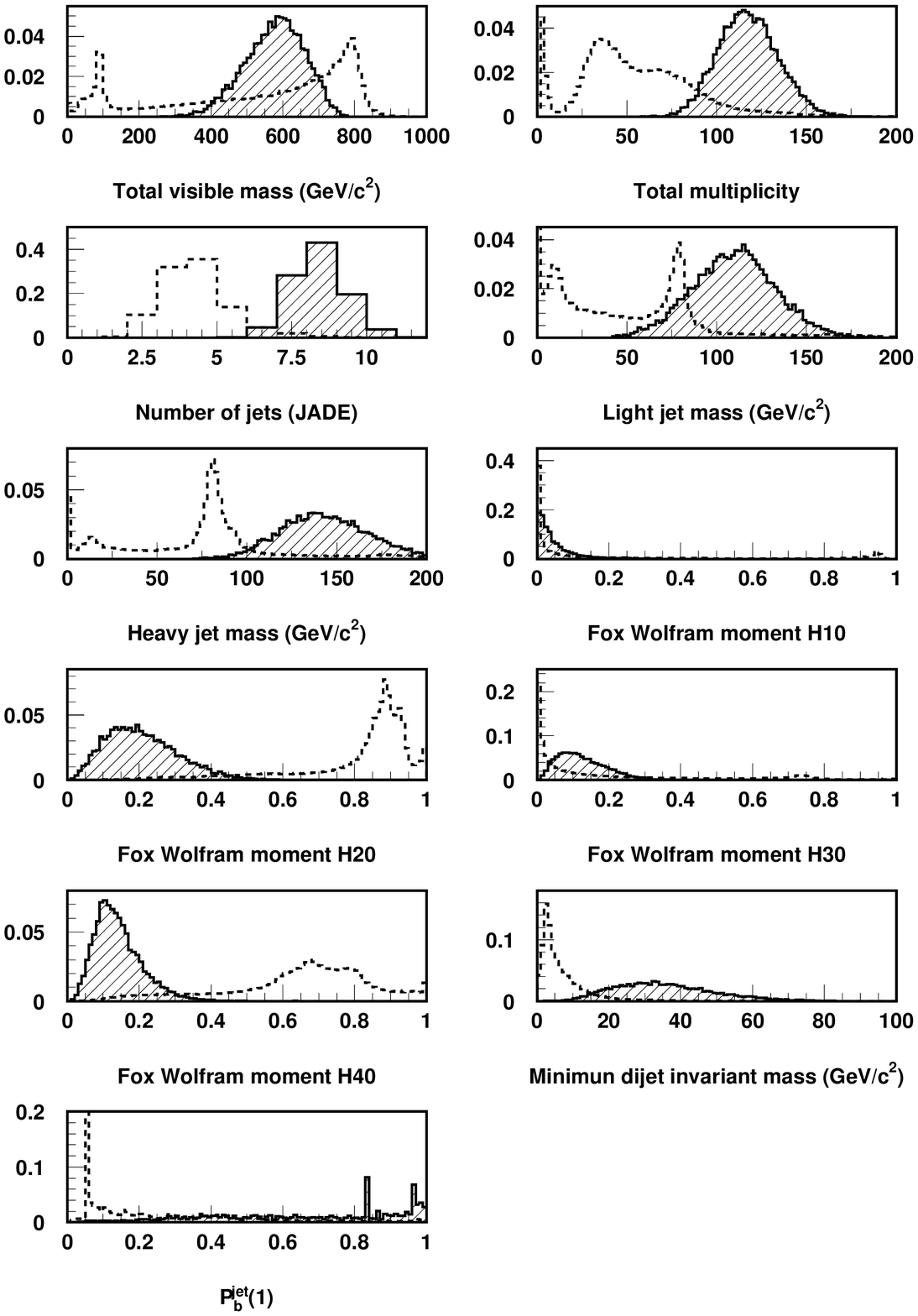,height=17cm,width=15cm}
\end{center}
\caption{\it The two like sign lepton plus 6 jet channel: selection variables (I). The signal (solid line) and the background (dashed line) are normalised to 1. The signal is shown for $M_{H} = 150$ GeV/c$^{2}$.} 
\label{hwwss1} 
\end{figure}

We first apply a set of criteria related to topological variables (figure~\ref{hwwss1}):

\begin{itemize}

\item 
400 GeV/c$^{2} <$ Total visible mass $<$ 700 GeV/c$^{2}$.
\item 
85 $<$ Total multiplicity $<$ 160.
\item 
Number of jets (including possible isolated leptons; JADE algorithm with $y_{cut}=1.10^{-3}$) $>$ 6.
\item 
Light jet mass $>$ 100 GeV/c$^{2}$.
\item 
Heavy jet mass $>$ 150 GeV/c$^{2}$.
\item 
Fox-Wolfram moment h10 $<$ 0.2.
\item 
Fox-Wolfram moment h20 $<$ 0.45.
\item 
Fox-Wolfram moment h30 $<$ 0.3.
\item 
Fox-Wolfram moment h40 $<$ 0.3.

\end{itemize}

The particles (discarding the charged leptons) of the surviving events are forced to a 6 jet configuration (using JADE) and the jets are flavour-tagged. The events are then required to fulfill the following criteria (figure~\ref{hwwss1}):

\begin{itemize}
\item 
Minimum dijet invariant mass $>$ 15 GeV/c$^{2}$.
\item 
$P^{jet}_{b}(1) >$ 0.2.

\end{itemize}

\begin{figure}
\begin{center}
\epsfig{figure=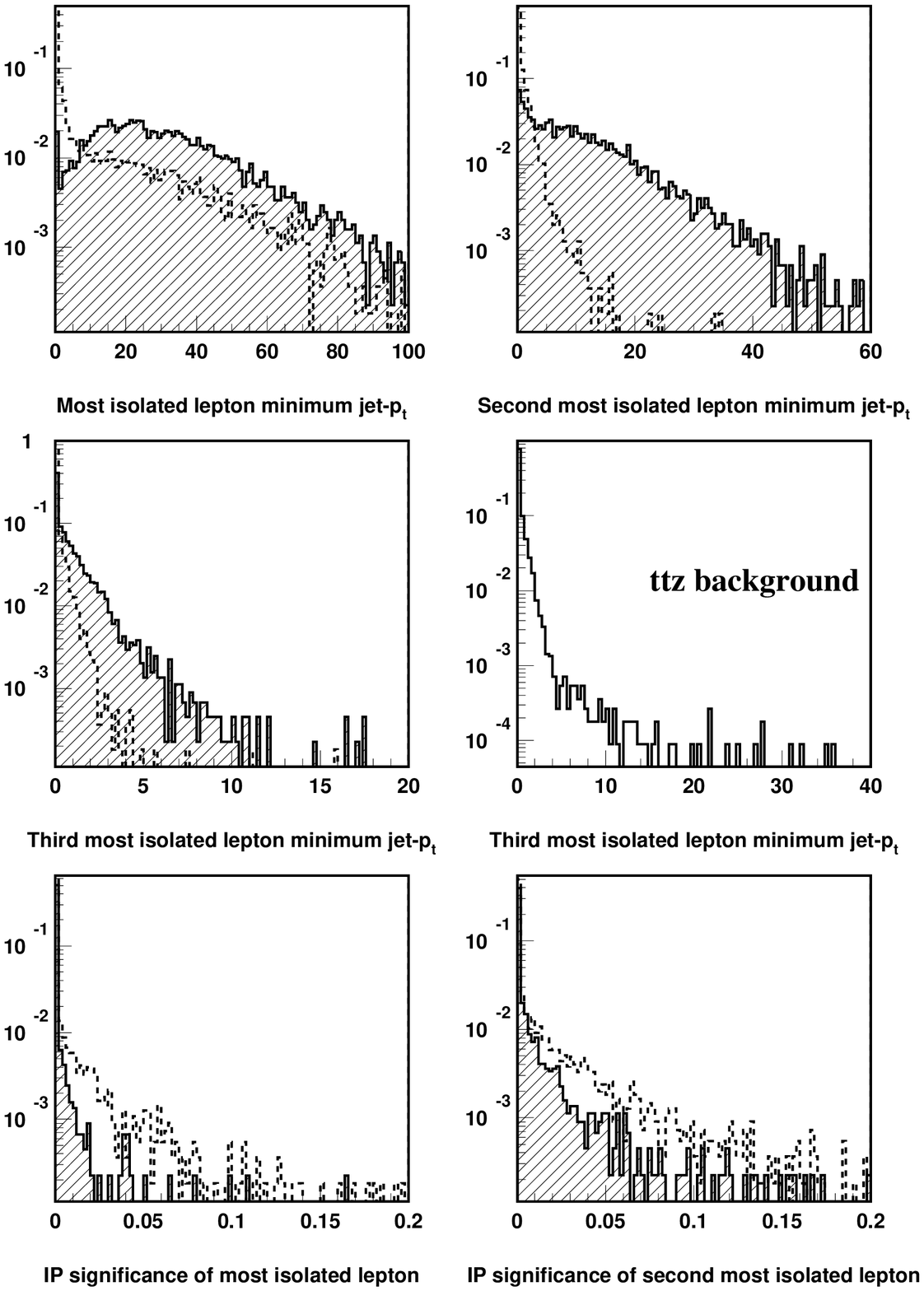,height=17cm,width=15cm}
\end{center}
\caption{\it The two like sign lepton plus 6 jet channel: selection variables (II). The signal (solid line) and the background (dashed line) are normalised to 1. The signal is shown for $M_{H} = 150$ GeV/c$^{2}$.} 
\label{hwwss2} 
\end{figure}

The request of only one b-jet allows to preserve a high selection efficiency. Now, we make use of the lepton content of the signal events to further eliminate the background. For each charged lepton reconstructed in the event, we calculate the transverse momentum with respect to the closest jet. Next, we classify the leptons from the most isolated to the least isolated according to this quantity. In the event, there should be 2 and only 2 isolated leptons. Moreover, they should have the same sign. As they come from the primary vertex, the significance of their impact parameter\footnote{For a charged track, the impact parameter is the distance of closest approach between this track and the primary vertex and the significance of the impact parameter is the ratio between the impact parameter and its estimated uncertainty.} should be small in comparison with the one of a lepton coming from a b- or c-meson decay. Therefore, we require (figure~\ref{hwwss2}):

\begin{itemize}
\item 
The 2 most isolated leptons have the same sign.
\item 
A $p_{t}$ between the most isolated lepton and the other jets $\ge$ 5 GeV/c.
\item 
A $p_{t}$ between the second most isolated lepton and the other jets $\ge$ 5 GeV/c.
\item 
A $p_{t}$ between the third most isolated lepton and the other jets $\le$ 5 GeV/c.
\item 
-0.001 $<$ Significance of most isolated lepton impact parameter $<$ 0.001.
\item 
-0.001 $<$ Significance of second most isolated lepton impact parameter $<$ 0.001.
\end{itemize}

The overall selection efficiencies and the corresponding effective cross-sections are shown in table~\ref{preseleffhww}. The main background is again due to top pair production. As the number of signal events is small in this channel but the purity high, the total uncertainty on $g_{ttH}$ is dominated by the statistical uncertainty. Therefore, the relative uncertainty on the residual background normalisation has only a modest influence on the total uncertainty. As a matter of fact, the procedure is not reoptimised for each value of the relative uncertainty on the residual background normalisation.

\begin{table}[h]
\centering
\begin{tabular}{||c||c||c|}
\hline
\hline
{ Final state} & {$\epsilon_{sel} $ ($\%$)} & {$\sigma _{eff} (fb)$}\\
\hline
{$t\overline tH \rightarrow 2l^{\pm}2\nu 2b4q$; $M_{H}=130$ GeV/c$^{2}$} & {$20.2$} & {1.26 $\cdot10^{-2}$}\\
{$t\overline tH \rightarrow 2l^{\pm}2\nu 2b4q$; $M_{H}=150$ GeV/c$^{2}$} & {$24.6$} & {2.72 $\cdot10^{-2}$}\\
{$t\overline tH \rightarrow 2l^{\pm}2\nu 2b4q$; $M_{H}=170$ GeV/c$^{2}$} & {$25.5$} & {2.96 $\cdot10^{-2}$}\\
{$t\overline tH \rightarrow 2l^{\pm}2\nu 2b4q$; $M_{H}=200$ GeV/c$^{2}$} & {$26.3$} & {1.47 $\cdot10^{-2}$}\\
\hline
{$ t\overline t$ } & {$4.40\cdot10^{-3}$} & {1.31$\cdot10^{-2}$}\\
{$ t\overline tZ$ } & {7.33$\cdot10^{-2}$} & {3.14$\cdot10^{-3}$}\\
{$ WW$ } & {$7.41\cdot10^{-5}$}  & {3.18$\cdot10^{-3}$}\\
{$ ZZ$ } & {$1.02\cdot10^{-3}$}  & {2.44$\cdot10^{-3}$}\\
{$ q\overline q$ } & {$<8.33\cdot10^{-5}$} & {$<$1.30$\cdot10^{-3}$}\\
{$ t\overline tH  (H \rightarrow WW) \nrightarrow 2l^{\pm}2\nu 2b4q$; $M_{H}=150$ GeV/c$^{2}$} & {$2.65\cdot10^{-1}$} & {2.77$\cdot10^{-3}$}\\
{$t\overline tH (H \nrightarrow WW)$; $M_{H}=150$ GeV/c$^{2}$}  & {8.0$\cdot10^{-2}$} & {3.94$\cdot10^{-4}$}\\
\hline
{Total background ; $M_{H}=150$ GeV/c$^{2}$} & {$3.84\cdot10^{-4}$} & {2.46$\cdot10^{-2}$}\\
\hline
\end{tabular}
\caption{\it The two like sign lepton plus 6 jet channel: selection efficiencies ($\epsilon_{sel} $) and corresponding effective cross-sections ($\sigma _{eff}$) for various Higgs boson masses. The 4 top lines stand for the selected signal final state, while the next lines break down the different background components (including those due to $t\overline tH$ events with another Higgs decay channel).}
\label{preseleffhww}
\end{table}

\subsubsection{Results}

The resolutions on the measurement of $g_{ttH}$ are shown in table~\ref{hwwsslepresults} and on figure \ref{plotresults}. They range from 12.8\% (13.4\%) to 24.9\% (26.3\%) for $\frac{\Delta \sigma _{BG}^{eff}}{\sigma _{BG}^{eff}} = 5\%$ (10\%).  For the lowest values of the Higgs mass range, the resolution is not good as the branching ratio of the Higgs into pairs of W's is very small. For $M_{H} \approx $ 170 GeV/c$^{2}$, the resolution curve has a minimum which corresponds to the maximum of the branching ratio. Then, for higher mass values, the resolution degrades due to the decrease of the cross section and of the branching ratio.

\begin{table}[h]
\centering
\begin{tabular}{||c||c||c||c||c||c||c|}
\hline
\hline
{ $M_{H}$ (GeV/c$^{2}$)} &{$\frac{\Delta \sigma _{BG}^{eff}}{\sigma _{BG}^{eff}}$} & {$\epsilon_{sel}^{signal} $} & {$\rho_{sel}^{sample}$} & {$\big(\frac{\Delta g_{ttH}}{g_{ttH}}\big)_{stat}$} & {$\big(\frac{\Delta g_{ttH}}{g_{ttH}}\big)_{syst}$} &  {$\frac{\Delta g_{ttH}}{g_{ttH}}$}\\
\hline
\hline
{ 130} &{$5\%$} &{$20.2\%$}&{$33.9\%$}&{$24.4\%$}&{$4.9\%$}&{$24.9\%$}\\
{    } &{$10\%$}&{$20.2\%$}&{$33.9\%$}&{$24.4\%$}&{$9.9\%$}&{$26.3\%$}\\
\hline
{ 150} &{$5\%$} &{$24.6\%$}&{$52.1\%$}&{$13.4\%$}&{$2.3\%$}&{$13.6\%$}\\
{    } &{$10\%$}&{$24.6\%$}&{$52.1\%$}&{$13.4\%$}&{$4.6\%$}&{$14.2\%$}\\
\hline
{ 170} &{$5\%$} &{$25.5\%$}&{$54.0\%$}&{$12.7\%$}&{$2.2\%$}&{$12.8\%$}\\
{    } &{$10\%$}&{$25.5\%$}&{$54.0\%$}&{$12.7\%$}&{$4.3\%$}&{$13.4\%$}\\
\hline
{ 200} &{$5\%$} &{$26.3\%$}&{$37.3\%$}&{$21.6\%$}&{$4.3\%$}&{$22.0\%$}\\
{    } &{$10\%$}&{$26.3\%$}&{$37.3\%$}&{$21.6\%$}&{$8.5\%$}&{$23.2\%$}\\
\hline
\hline
\end{tabular}
\caption{\it The two like sign lepton plus 6 jet channel: expected uncertainty on the measurement of $g_{ttH}$ for various Higgs boson masses. Selection efficiency of the signal ($\epsilon_{sel}^{signal} $) and purity of the selected sample ($\rho_{sel}^{sample}$) are also shown.}
\label{hwwsslepresults}
\end{table}

\subsubsection{Precision loss due to 6 fermion processes}

The 6 fermion processes which may degrade the resolution on $g_{ttH}$ in this channel are listed in table~\ref{seleffhwwss6f} with their cross section and selection efficiency for $M_{H}=150$ GeV/c$^{2}$. The loss of resolution on $g_{ttH}$ when the 6 fermion background is included in the analysis is shown in table~\ref{hwwssresults6f}, showing that it may be neglected.

\begin{table}[h]
\centering
\begin{tabular}{||c|c|c||}
\hline
{ Final state}& {$\sigma  (fb)$} &{$\epsilon_{sel} $}\\
\hline
\hline
{$b\overline bu\overline de^{-}\overline \nu_{e} $ $^{*}$ } & {$\sim 1.5$}  & {$8.4\cdot 10^{-3}\%$} \\
\hline
{$b\overline bt\overline b d\overline u$ $^{*}$ } & {$\sim 6\cdot10^{-3}$}  & {$<0.2\%$} \\
{$b\overline bt\overline b \mu\overline \nu$  $^{*}$} & {$\sim 1.5\cdot10^{-3}$}  & {$0.4\%$} \\
\hline
{$t\overline tq\overline qq'\overline q'$ $(q,q'=u,d,s)$ } & {$1.2\cdot10^{-2}$}  & {$0.2\%$}\\
{$q\overline qt\overline b l\overline \nu_{l}$ $(q=u,d,s,c$; $l=e^-,\mu^-)$ $^{*}$} & {$\sim 7.6\cdot10^{-2}$}  &
 {$0.2\%$} \\
{$q\overline qt\overline b d\overline u$ $(q=u,d,s,c)$ $^{*}$} & {$\sim 1.3\cdot10^{-1}$}  & {$0.05\%$} \\
\hline
\hline
\end{tabular}
\caption{\it The two like sign lepton plus 6 jet channel: cross section ($\sigma$) and selection efficiencies ($\epsilon_{sel} $) of the 6 fermion processes for $\frac{\Delta \sigma _{BG}^{eff}}{\sigma _{BG}^{eff}}$ = $5\%$  and $M_{H}=150$ GeV/c$^{2}$. For events marked with a $^{*}$, some diagrams were removed from the calculation to avoid double counting.}
\label{seleffhwwss6f}
\end{table}

\vspace{-1cm}

\begin{table}[h!]
\centering
\begin{tabular}{||c||c||c|}
\hline
{$\frac{\Delta \sigma _{BG}^{eff}}{\sigma _{BG}^{eff}}$}  &  {$\frac{\Delta g_{ttH}}{g_{ttH}}$}(without 6f) &  {$\frac{\Delta g_{ttH}}{g_{ttH}}$(with 6f)}\\
\hline
\hline
{$5\%$} & {$ 13.6\%$} & {$ 13.9\%$}\\
{$10\%$} &{$ 14.2\%$} & {$ 14.5\%$}\\
\hline
\end{tabular}
\caption{\it The two like sign lepton plus 6 jet channel: expected uncertainty on the measurement of $g_{ttH}$ without and with inclusion of the 6 fermion processes in the analysis for $M_{H}=150$ GeV/c$^{2}$.}
\label{hwwssresults6f}
\end{table}

\subsection{The single lepton plus 8 jet channel}

\subsubsection{Introduction}

The final state follows from the process:
\\

$e^{+}e^{-} \rightarrow t\overline tH \rightarrow W^{+}bW^{-}\overline bW^{+}W^{-} \rightarrow l\nu 2b 6q$\\

Its branching ratio is:

\begin{eqnarray}
BR(t\overline tH \rightarrow l\nu 2b6q) & = & 4*BR(H \rightarrow W^{+}W^{-}) *BR(W \rightarrow l\nu)*(BR(W \rightarrow 2q))^{3}\nonumber\\
& \approx &  40\%*BR(H \rightarrow W^{+}W^{-}) \nonumber
\end{eqnarray}

This channel is thus characterized by a missing 4-momentum, one prompt charged lepton, 6 light quark jets and 2 b-jets. This signature is less singular than the one of the previous channel but the branching ratio is about 4 times larger. This final state is close to the one of the $H \rightarrow b\overline b$ semileptonic channel, the analysis will thus be very similar. Notice that, here, $ t\overline tH (H \rightarrow WW) \nrightarrow l\nu 2b6q$ and $ t\overline tH (H \nrightarrow WW)$ act as background processes.
\subsubsection{Sequential analysis}

\begin{figure}
\begin{center}
\epsfig{figure=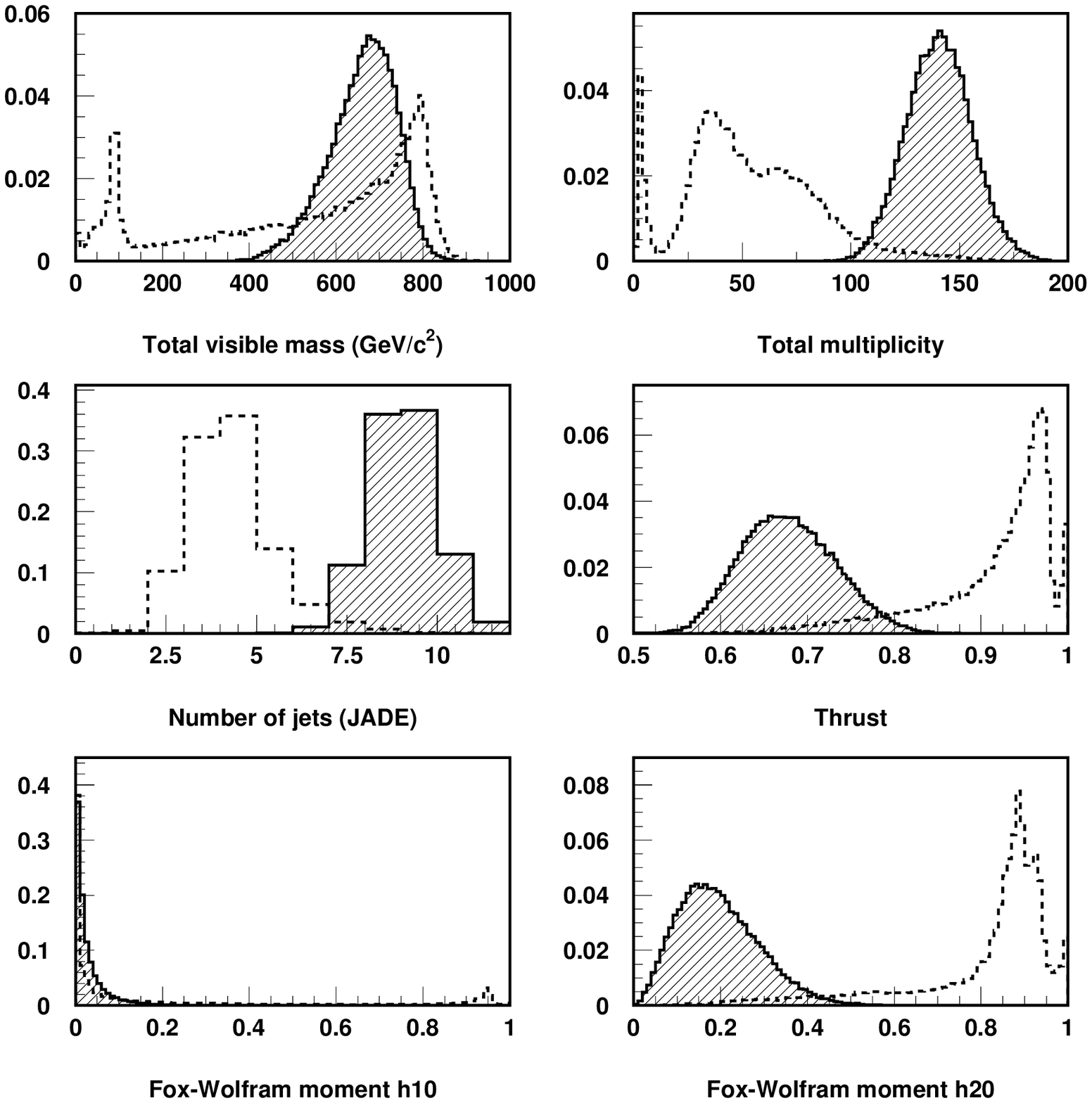,height=17cm,width=15cm}
\end{center}
\caption{\it The single lepton plus 8 jet channel: preselection variables (I). The signal (solid line) and the background (dashed line) are normalised to 1. The signal is shown for $M_{H} = 150$ GeV/c$^{2}$.}
\label{hww_onelep_1} 
\end{figure}

\begin{figure}
\begin{center}
\epsfig{figure=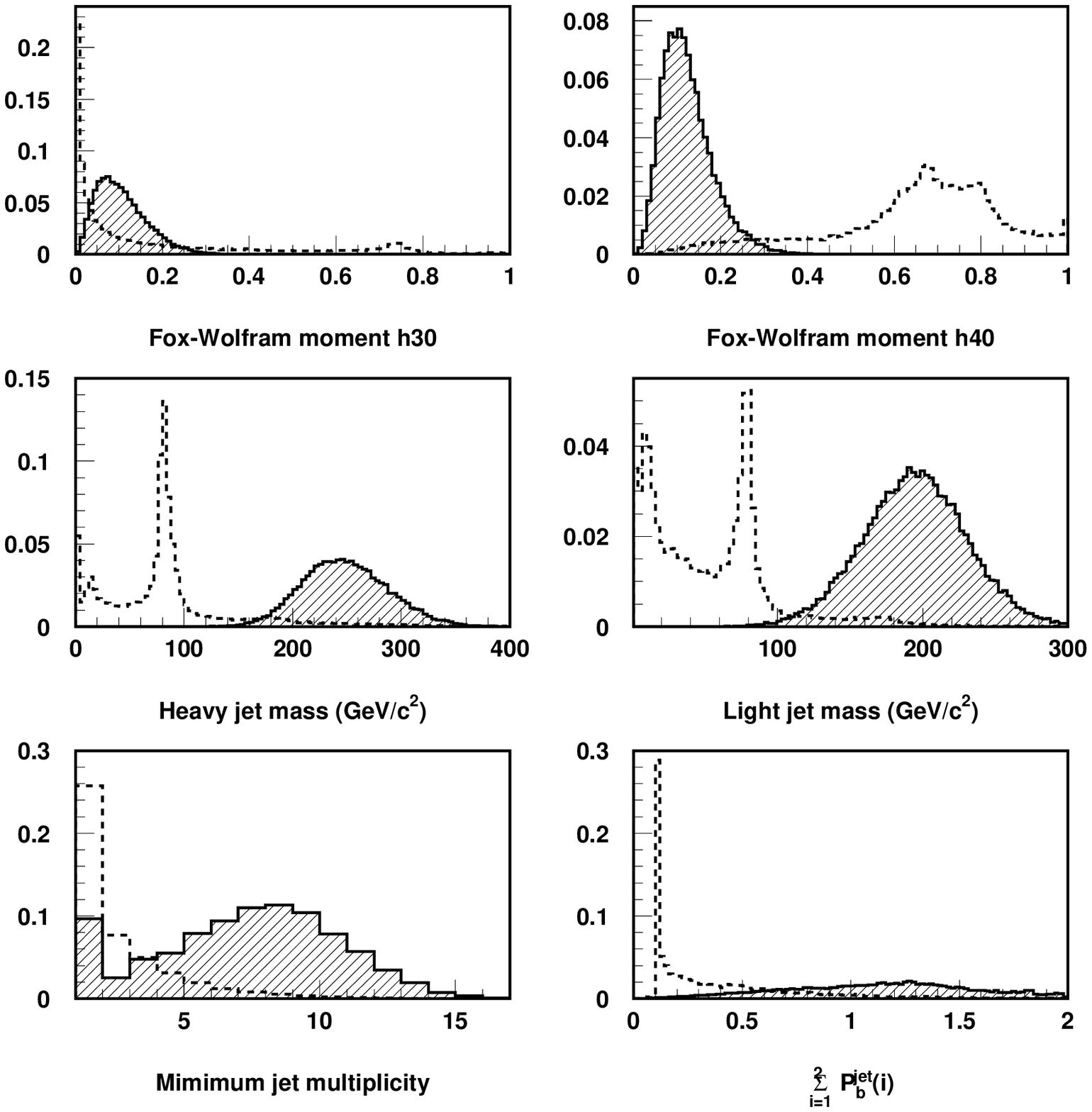,height=17cm,width=15cm}
\end{center}
\caption{\it The single lepton plus 8 jet channel: preselection variables (II). The signal (solid line) and the background (dashed line) are normalised to 1. The signal is shown for $M_{H} = 150$ GeV/c$^{2}$.}
\label{hww_onelep_2} 
\end{figure}

We first request the presence in the event of at least one charged lepton (a $\mu^{\pm}$ or a $e^{\pm}$). Then, we require (figure~\ref{hww_onelep_1} and~\ref{hww_onelep_2}):

\begin{itemize}
 \item 
500 GeV/c$^{2}$$ < $ Total visible mass $ < $ 750 GeV/c$^{2}$.
 \item 
Total multiplicity $ \ge $ 110.
\item 
Number of jets (including possible isolated leptons; JADE algorithm with $y_{cut}=1.10^{-3}$) $\ge$ 8.
\item 
Thrust $\le$ 0.8
\item 
Fox-Wolfram moment h10 $\le$ 0.1.
\item 
Fox-Wolfram moment h20 $\le$ 0.5.
\item 
Fox-Wolfram moment h30 $\le$ 0.3.
\item 
Fox-Wolfram moment h40 $\le$  0.3.
\item 
Light jet mass $\ge$ 100 GeV/c$^{2}$.
\item 
Heavy jet mass $\ge$ 150 GeV/c$^{2}$.

\end{itemize}

To identify an energetic and isolated charged lepton, we proceed as in the $H \rightarrow b\overline b$ semileptonic channel. The only difference is that, here, to tag the lepton as well as when it has been tagged, the remaining particles are forced to a 8 jet configuration (with the JADE algorithm). Then, the jets are flavour-tagged and finally, we require (figure~\ref{hww_onelep_2}):

\begin{itemize}
\item 
Mininum jet multiplicity $\ge$ 3.
 \item 
$\sum_{i=1}^{2} P^{jet}_{b}(i) \ge$ 0.2.
\end{itemize}

The overall preselection efficiencies and the corresponding effective cross-sections are shown in table~\ref{preseleffhwwonelep}. The main background after this preselection comes from the $t\overline t$ production.

\begin{table}[h]
\centering
\begin{tabular}{||c||c||c|}
\hline
\hline
{ Final state} & {$\epsilon_{presel}$ ($\%$)} & {$\sigma _{eff} (fb)$}\\
\hline
{$t\overline tH \rightarrow l\nu 2b6q$; $M_{H}=130$ GeV/c$^{2}$} & {$50.3$} & {1.31$\cdot10^{-1}$}\\
{$t\overline tH \rightarrow l\nu 2b6q$; $M_{H}=150$ GeV/c$^{2}$} & {$55.6$} & {2.57$\cdot10^{-1}$}\\
{$t\overline tH \rightarrow l\nu 2b6q$; $M_{H}=170$ GeV/c$^{2}$} & {$58.3$} & {2.83$\cdot10^{-1}$}\\
{$t\overline tH \rightarrow l\nu 2b6q$; $M_{H}=200$ GeV/c$^{2}$} & {$60.8$} & {1.42$\cdot10^{-1}$}\\
\hline
{$ t\overline t$ } & {$2.76$} & {8.22}\\
{$ t\overline tZ$ } & {$13.8$} & {5.91$\cdot10^{-1}$}\\
{$ WW$ } & {1.73$\cdot10^{-2}$} & {7.45$\cdot10^{-1}$}\\
{$ ZZ$ } & {6.55$\cdot10^{-2}$} & {1.60$\cdot10^{-1}$}\\
{$ q\overline q$ } & {$2.33\cdot10^{-3}$} & {3.63$\cdot10^{-2}$}\\
{$ t\overline tH (H \rightarrow WW)\nrightarrow l\nu 2b6q$; $M_{H}=150$ GeV/c$^{2}$} & {$18.9$} & {1.31$\cdot10^{-1}$}\\
{$ t\overline tH (H \nrightarrow WW)$; $M_{H}=150$ GeV/c$^{2}$}  & {$26.5$} & {1.30$\cdot10^{-1}$}\\
\hline
{Total background; $M_{H}=150$ GeV/c$^{2}$} & {$1.54\cdot10^{-1}$} & {10.0}\\
\hline
\end{tabular}
\caption{\it The single lepton plus 8 jet channel: preselection efficiencies ($\epsilon_{presel} $) and corresponding effective cross-sections ($\sigma _{eff}$) for various Higgs boson masses. The 4 top lines stand for the selected signal final state, while the next lines break down the different background components (including those due to $t\overline tH$ events with another Higgs decay channel)}
\label{preseleffhwwonelep}
\end{table}

\subsubsection{Neural network analysis}

The distributions of some of the preselection variables still contain unused information. They are recycled to train a neural network together with other variables related to b-tagging, to the prompt lepton and to the missing momentum.

\begin{figure}
\begin{center}
\epsfig{figure=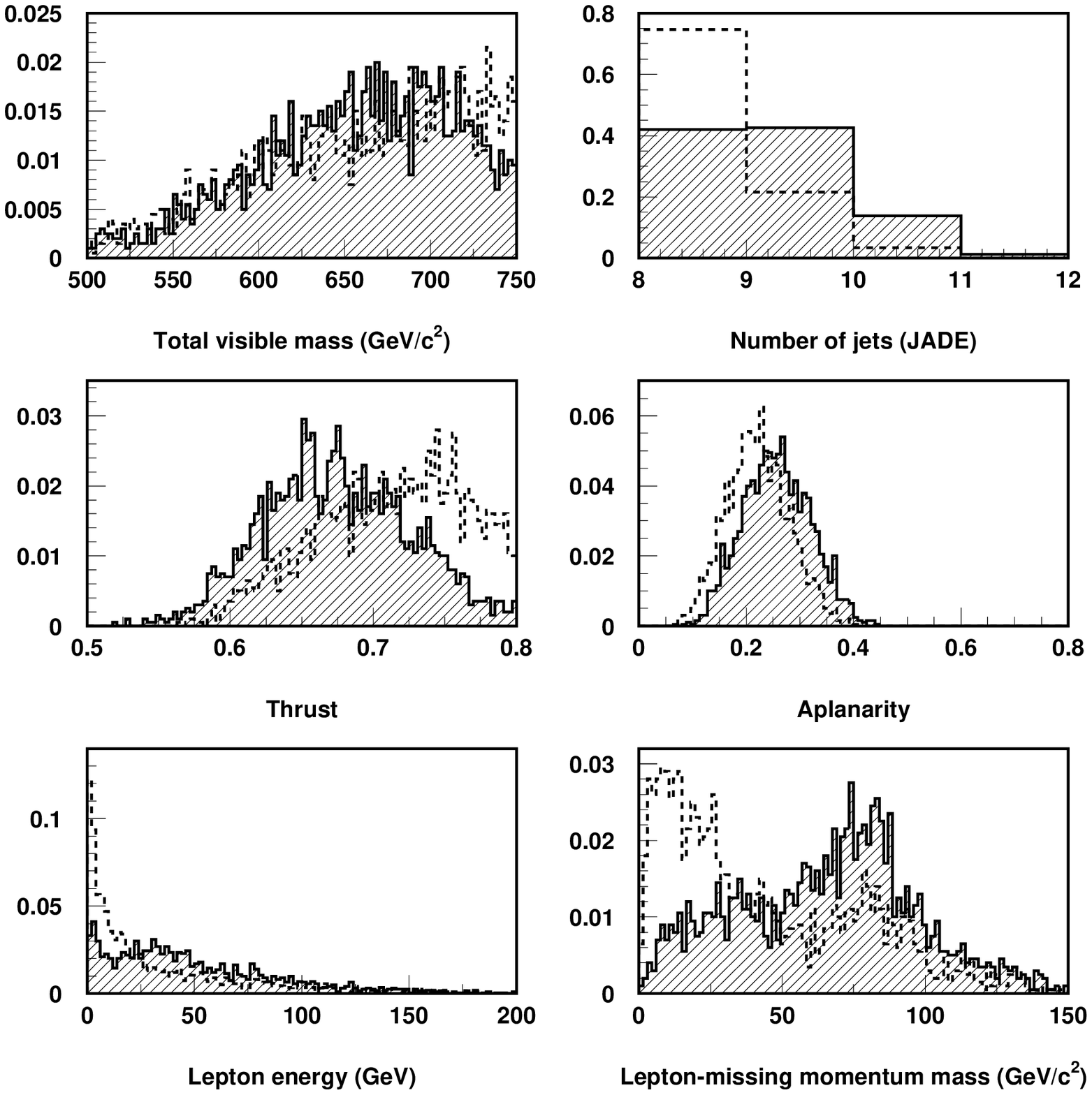,height=17cm,width=15cm}
\end{center}
\caption{\it The single lepton plus 8 jet channel: variables (after preselection) used for the neural network analysis (I). The signal (solid line) and the background (dashed line) are normalised to 1. The signal is shown for $M_{H} = 150$ GeV/c$^{2}$.}
\label{hww_onelep_3} 
\end{figure}

\begin{figure}
\begin{center}
\epsfig{figure=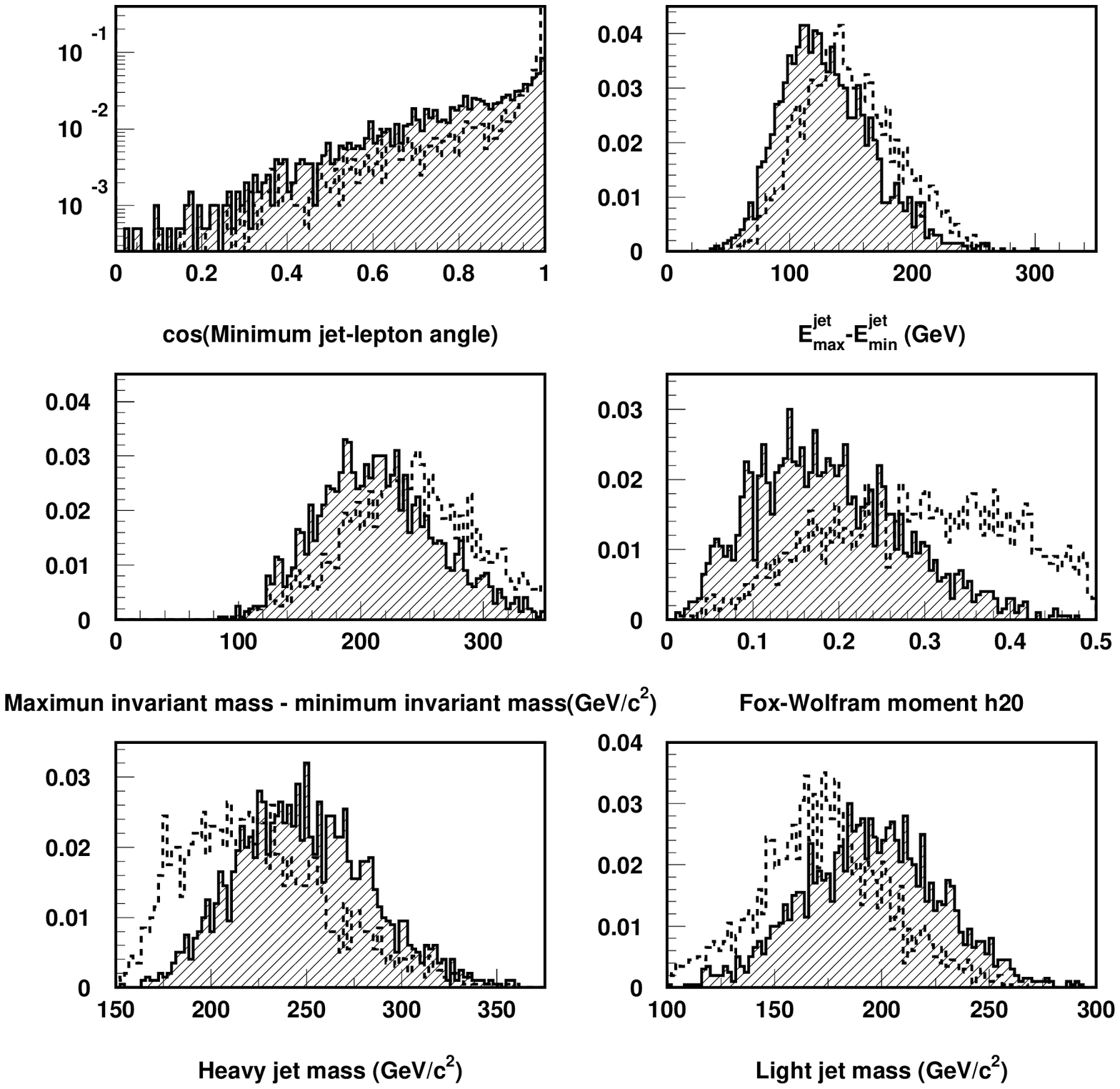,height=17cm,width=15cm}
\end{center}
\caption{\it The single lepton plus 8 jet channel: variables (after preselection) used for the neural network analysis (II). The signal (solid line) and the background (dashed line) are normalised to 1. The signal is shown for $M_{H} = 150$ GeV/c$^{2}$.}
\label{hww_onelep_4} 
\end{figure}

The complete list of variables used in the network is (figure~\ref{hww_onelep_3} and~\ref{hww_onelep_4}):
\begin{itemize}
\item 
The total visible mass.
\item 
The number of jets (including possible isolated leptons; JADE algorithm with $y_{cut}=1.10^{-3}$).
\item 
The thrust.
\item 
The aplanarity.
\item 
The energy of the tagged lepton.
\item 
The invariant mass of the system made of the lepton and the missing momentum.
\item 
The cosine of the angle between the tagged lepton and the closest jet directions.
\item 
$E^{jet}_{max}-E^{jet}_{min}$.
\item 
Maximum dijet invariant mass$ - $minimum dijet invariant mass.
\item 
The second Fox-Wolfram moment h20.
\item 
The light jet mass.
\item 
The heavy jet mass.

\end{itemize}

The lepton related variables are calculated once the event has been forced to the 1 lepton plus 8 jet configuration, as well as the variables $E^{jet}_{max}$, $E^{jet}_{min}$, maximum dijet invariant mass and minimum dijet invariant mass.

\subsubsection{Results}

\begin{table}[h]
\centering
\begin{tabular}{||c||c||c|}
\hline
\hline
{ Final state} &{$\epsilon_{sel} $ ($\%$)}& {$\sigma _{eff} (fb)$}\\
\hline
{$t\overline tH \rightarrow l\nu 2b6q$} & {$16.7$} & {7.71$\cdot10^{-2}$}\\
\hline
{$ t\overline t$ } & {5.65$\cdot10^{-2}$} & {1.68$\cdot10^{-1}$}\\
{$ t\overline tZ$ } & {4.77$\cdot10^{-1}$} & {2.04$\cdot10^{-2}$}\\
{$ WW$ } & {$1.48\cdot10^{-4}$} & {6.37$\cdot10^{-3}$}\\
{$ ZZ$ } & {$3.02\cdot10^{-3}$} & {7.25$\cdot10^{-3}$}\\
{$ q\overline q$ } & {$<10^{-4}$} & {$<1.6\cdot10^{-3}$}\\
{$t\overline tH (H\rightarrow WW)\nrightarrow l\nu 2b6q$} & {$1.24$} & {8.59$\cdot10^{-3}$}\\
{$ t\overline tH (H \nrightarrow WW)$}  & {$2.35$} & {1.16$\cdot10^{-2}$}\\
\hline
{Total background } & {$3.30\cdot10^{-3}$} & {2.22$\cdot10^{-1}$}\\
\hline
\end{tabular}
\caption{\it The single lepton plus 8 jet channel: selection efficiencies ($\epsilon_{sel} $) and corresponding effective cross-sections ($\sigma _{eff}$) for $\frac{\Delta \sigma _{BG}^{eff}}{\sigma _{BG}^{eff}}$ = $5\%$  and $M_{H}=150$ GeV/c$^{2}$.}
\label{seleffhwwonelep}
\end{table}

\begin{figure}
\begin{center}
\epsfig{figure=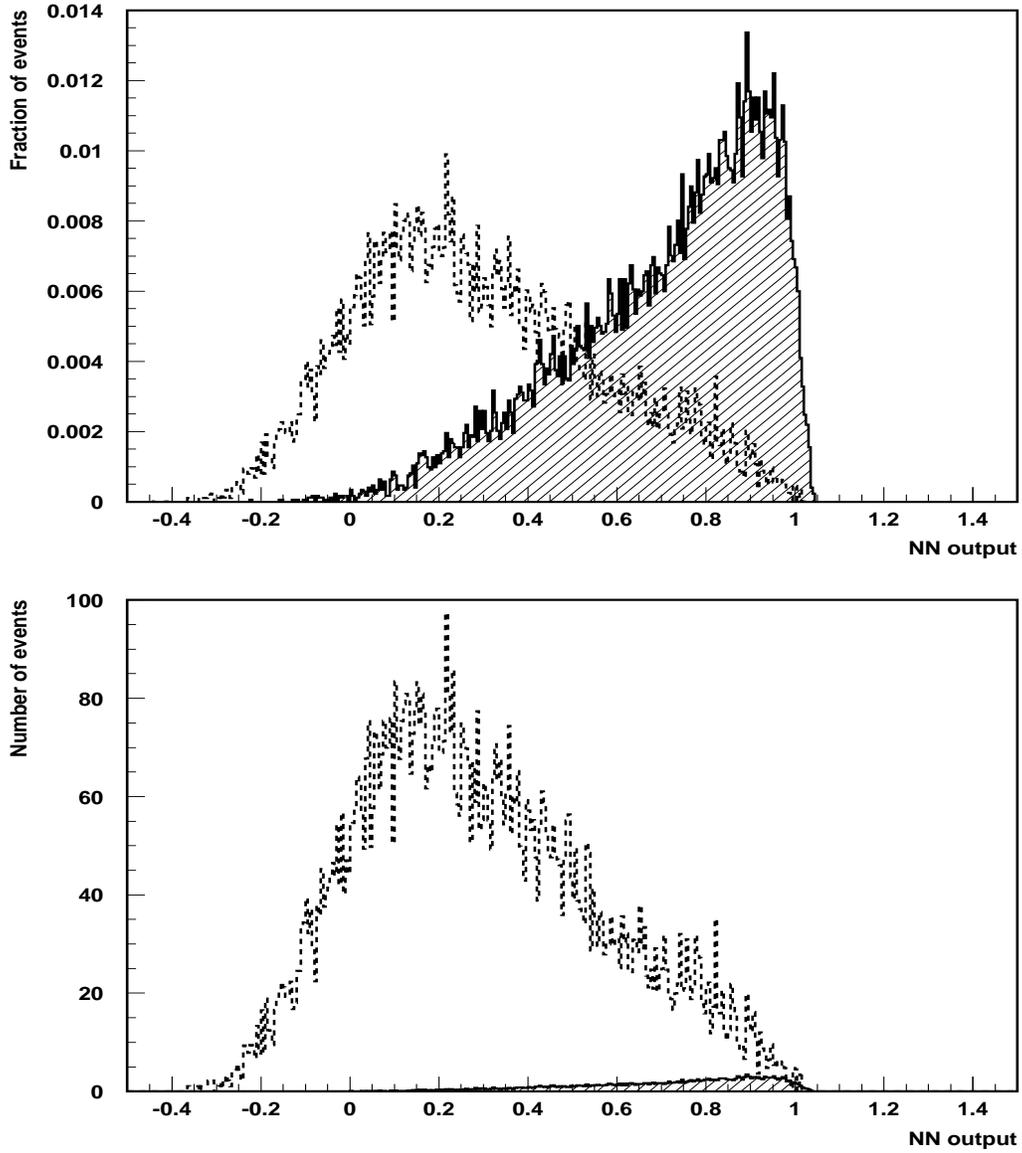,height=17cm,width=15cm}
\end{center}
\caption{\it The single lepton plus 8 jet channel: neural network output. Top, the signal (solid line) and the background (dashed line) are normalised to 1. Bottom, the signal (solid line) and the background (dashed line) are normalised to the expected number of events. The signal is shown for $M_{H} = 150$ GeV/c$^{2}$.}
\label{hww_onelep_5} 
\end{figure}

The result of the training session is shown in figure~\ref{hww_onelep_5}. We apply the same procedure as for the $H \rightarrow b\overline b$ semileptonic channel (see section 5.2.4) and an example of the values of the selection efficiencies and the corresponding effective cross-sections are shown in table~\ref{seleffhwwonelep}. The main background after selection is again due to top-pair production. The precisions we get for $g_{ttH}$ are shown in table~\ref{hwwonelepresults} and on figure \ref{plotresults}. They range from 10.2\% (12.1\%) to 35.2\% (47.2\%) for $\frac{\Delta \sigma _{BG}^{eff}}{\sigma _{BG}^{eff}} = 5\%$ (10\%). The remarks made for the previous channel apply here as well. For the lowest values of the Higgs mass range, the resolution is not good as the branching ratio of the Higgs into pairs of W's is very small. For $M_{H} \approx $ 170 GeV/c$^{2}$, the resolution curve has a minimum which corresponds to the maximum of the branching ratio. Then, for higher mass values, the resolution degrades due to the decrease of the cross section and of the branching ratio.

\begin{table}[h]
\centering
\begin{tabular}{||c||c||c||c||c||c||c|}
\hline
\hline
{ $M_{H}$ (GeV/c$^{2}$)} &{$\frac{\Delta \sigma _{BG}^{eff}}{\sigma _{BG}^{eff}}$} & {$\epsilon_{sel}^{signal} $} & {$\rho_{sel}^{sample}$} & {$\big(\frac{\Delta g_{ttH}}{g_{ttH}}\big)_{stat}$} & {$\big(\frac{\Delta g_{ttH}}{g_{ttH}}\big)_{syst}$} &  {$\frac{\Delta g_{ttH}}{g_{ttH}}$}\\
\hline
\hline
{ 130} &{$5\%$} &{$8.9\%$}&{$12.2\%$}&{$30.1\%$}&{$18.2\%$}&{$35.2\%$}\\
{    } &{$10\%$} &{$8.9\%$}&{$12.2\%$}&{$30.1\%$}&{$36.4\%$}&{$47.2\%$}\\
\hline
{ 150} &{$5\%$} &{$16.7\%$}&{$25.8\%$}&{$11.3\%$}&{$7.3\%$}&{$13.5\%$}\\
{    } &{$10\%$}&{$9.3\%$}&{$33.4\%$}&{$13.4\%$}&{$10.1\%$}&{$16.8\%$}\\
\hline
{ 170} &{$5\%$} &{$15.9\%$}&{$38.0\%$}&{$9.3\%$}&{$4.1\%$}&{$10.2\%$}\\
{    } &{$10\%$}&{$12.5\%$}&{$42.5\%$}&{$10.0\%$}&{$6.8\%$}&{$12.1\%$}\\
\hline
{ 200} &{$5\%$} &{$16.0\%$}&{$27.5\%$}&{$15.7\%$}&{$6.6\%$}&{$17.1\%$}\\
{    } &{$10\%$}&{$12.5\%$}&{$32.2\%$}&{$16.5\%$}&{$10.6\%$}&{$19.6\%$}\\
\hline
\hline
\end{tabular}
\caption{\it The single lepton plus 8 jet channel: expected uncertainty on the measurement of $g_{ttH}$ for various Higgs boson masses. Selection efficiency of the signal ($\epsilon_{sel}^{signal} $) and purity of the selected sample ($\rho_{sel}^{sample}$) are also shown.}
\label{hwwonelepresults}
\end{table}

\subsubsection{Precision loss due to 6 fermion processes}

\begin{table}[!]
\centering
\begin{tabular}{||c|c|c||}
\hline
{ Final state}& {$\sigma  (fb)$} &{$\epsilon_{sel} $}\\
\hline
\hline
{$b\overline bu\overline de^{-}\overline \nu_{e} $  $^{*}$} & {$\sim 1.5$}  & {$0.04\%$} \\
\hline
{$b\overline bt\overline b d\overline u$  $^{*}$} & {$\sim 6\cdot10^{-3}$}  & {$2.4\%$} \\
{$b\overline bt\overline b \mu\overline \nu$  $^{*}$} & {$\sim 1.5\cdot10^{-3}$}  & {$3.4\%$} \\
\hline
{$t\overline tq\overline qq'\overline q' $ $(q,q'=u,d,s)$ } & {$1.2\cdot10^{-2}$}  & {$7.6\%$}\\
{$q\overline qt\overline b l\overline \nu$ $(q=u,d,s,c;$ $l=e^-,\mu^-$) $^{*}$ } & {$\sim 7.6\cdot10^{-2}$}  & {$
1.7\%$} \\
{$q\overline qt\overline b d\overline u$ $(q=u,d,s,c)$ $^{*}$} & {$\sim 1.3\cdot10^{-1}$}  & {$1.2\%$} \\
\hline
\end{tabular}
\caption{\it The single lepton plus 8 jet channel: cross section ($\sigma$) and selection efficiencies ($\epsilon_{sel} $) of the 6 fermion processes for $\frac{\Delta \sigma _{BG}^{eff}}{\sigma _{BG}^{eff}}$ = $5\%$  and $M_{H}=150$ GeV/c$^{2}$. For events marked with a $^{*}$, some diagrams were removed from the calculation to avoid double counting.}
\label{seleffhwwonelep6f}
\end{table}

The 6 fermion processes which may degrade the resolution on $g_{ttH}$ in this channel are the same than in the previous channel. They are listed in table~\ref{seleffhwwonelep6f} with their cross section and selection efficiency for $M_{H}=150$ GeV/c$^{2}$. The loss of resolution on $g_{ttH}$ when the 6 fermion background is included in the analysis is shown in table~\ref{hwwonelepresults6f}. This channel is slightly more affected by the 6 fermion background than the previous ones because its signature is less distinctive. The loss of resolution is however still very small.

\begin{table}[h!]
\centering
\begin{tabular}{||c||c||c|}
\hline
{$\frac{\Delta \sigma _{BG}^{eff}}{\sigma _{BG}^{eff}}$}  &  {$\frac{\Delta g_{ttH}}{g_{ttH}}$}(without 6f) &  {$\frac{\Delta g_{ttH}}{g_{ttH}}$(with 6f)}\\
\hline
\hline
{$5\%$} &{$ 13.5 \%$} & {$\sim 14 \%$}\\
{$10\%$} &{$ 16.8 \%$} & {$\sim 17.5 \%$}\\
\hline
\end{tabular}
\caption{\it The single lepton plus 8 jet channel: expected uncertainty on the measurement of $g_{ttH}$ without and with inclusion of the 6 fermion processes in the analysis for $M_{H}=150$ GeV/c$^{2}$.}
\label{hwwonelepresults6f}
\end{table}

\section{Influence of the limited knowledge on the SM input parameters}

The results presented in this note were obtained for fixed values of the SM input parameters. The question was addressed whether the present limited knowledge on these values has any significant influence on the expected precision on $g_{ttH}$. 

The largest effect, if any, is expected from the present experimental uncertainty on the top quark mass (which is of the order of  5 GeV/c$^2$). A change in $m_t$ modifies several ingredients of the analysis: the selection efficiencies for signal and background, the top-pair production cross-section ($\sigma_{t\overline t}$), the signal cross-section ($\sigma_{t\overline t H}$) and the function $F(M_{H},m_{t},s)$ entering equation \ref{eq1} and \ref{eq2}, a.s.o. 

\begin{figure}
\begin{center}
\includegraphics[angle=0,width=15cm,height=16cm]{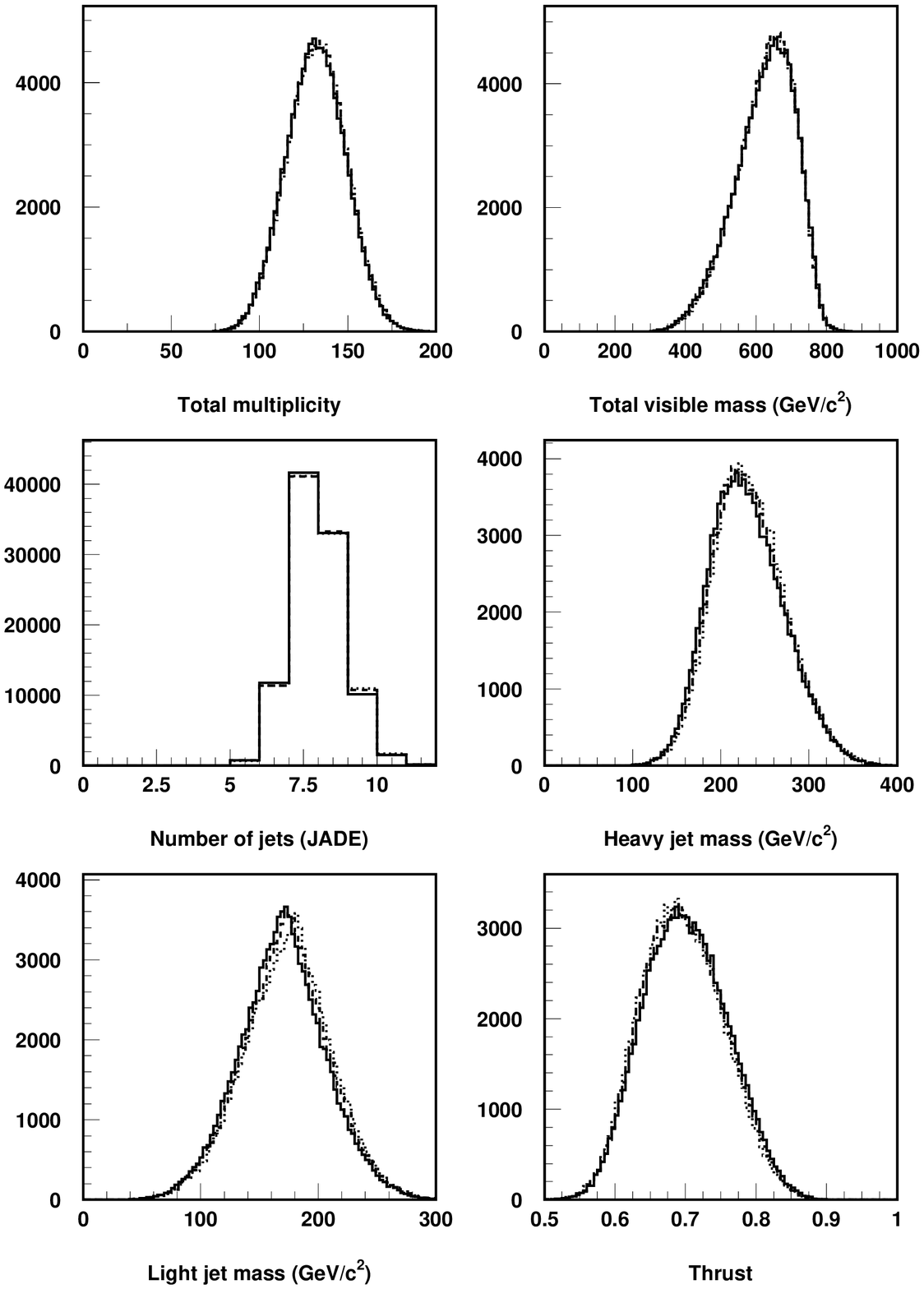}
\end{center}
\caption{\it The $H \rightarrow b \overline b$ semileptonic channel:  variables used in the analysis for three different top mass values: 170, 175 and 180 GeV/c$^{2}$ ($M_{H} = 120$ GeV/c$^{2}$) (I).}
\label{effect_mtop1} 
\end{figure}

\begin{figure}
\begin{center}
\includegraphics[angle=0,width=15cm,height=16cm]{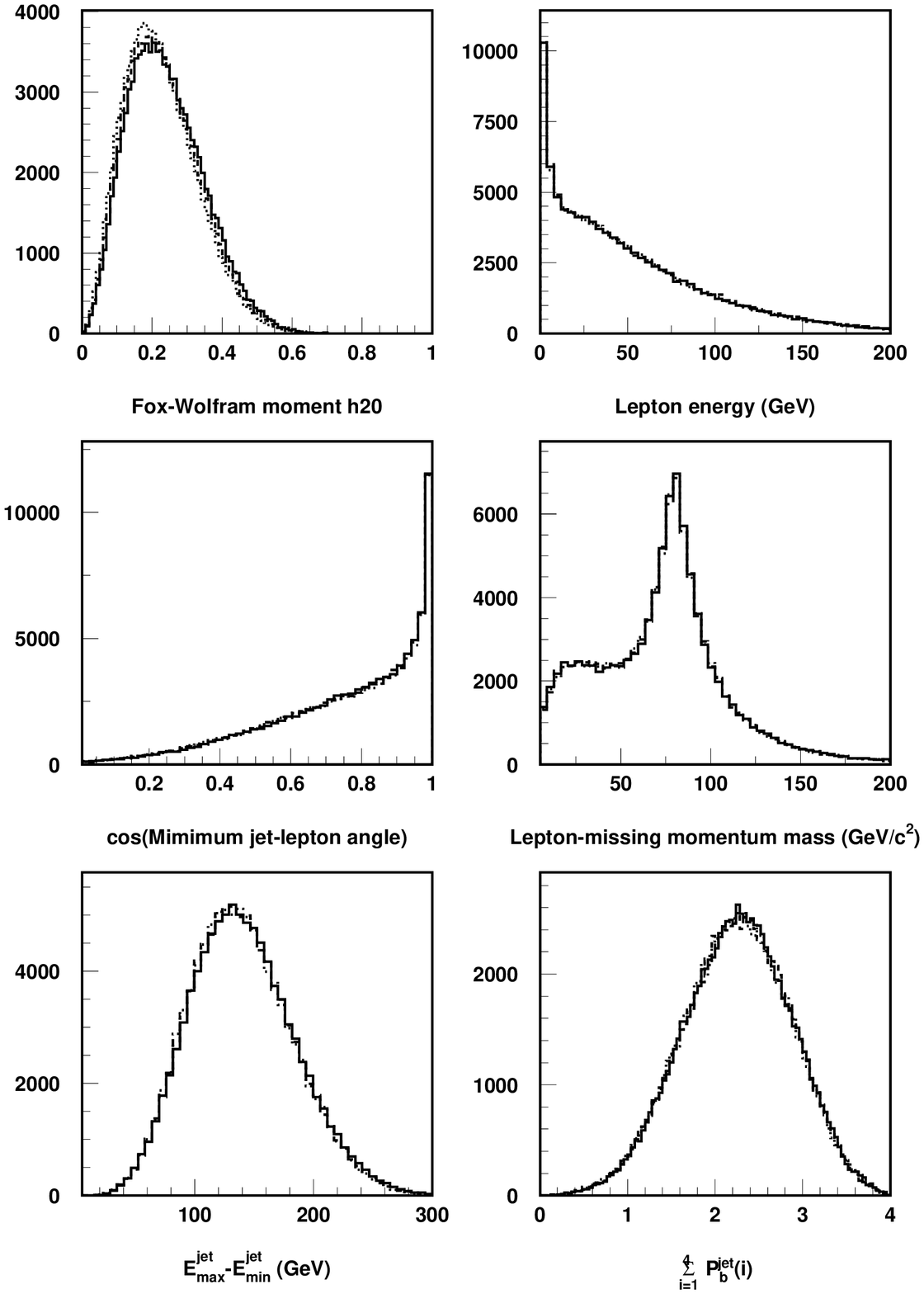}
\end{center}
\caption{\it The $H \rightarrow b \overline b$ semileptonic channel:  variables used in the analysis for three different top mass values: 170, 175 and 180 GeV/c$^{2}$ ($M_{H} = 120$ GeV/c$^{2}$) (II).}
\label{effect_mtop2} 
\end{figure}

The study presented here was performed by moving $m_t$ from its central value of 175 GeV/c$^2$ to 170 and 180 GeV/c$^2$. The change in $\sigma_{t\overline t H}$ was observed to be largest for low values of $M_H$ (i.e. 120 GeV/c$^2$), where it amounted to $\pm$ 5 \%, while the main background cross-section ($\sigma_{t\overline t}$) varied by $\pm$ 1 \% only. Signal events were generated with $m_t$ = 170 and 180 GeV/c$^2$, $M_H$ being fixed to 120 GeV/c$^2$, and the analysis of the Higgs decay into b-quark pairs was repeated for the so-called semileptonic channel (see section \ref{semilep}). The distributions of several variables entering the event selection are displayed on figure \ref{effect_mtop1} and \ref{effect_mtop2} for each value of $m_t$. The changes observed are tiny, illustrating that the preselection and selection efficiencies remain essentially unchanged. Overall, the study demonstrated that the precision on $g_{ttH}$ varies by a negligible amount when moving $m_t$ from 170 to 180 GeV/c$^2$. The conclusion is therefore that the present accuracy on the SM input parameters has no significant influence on the predicted accuracy on $g_{ttH}$.

\section{Conclusion and outlook}

The process $e^{+}e^{-} \rightarrow t\overline tH$ allows in principle a direct measurement of the top-Higgs Yukawa coupling. We presented a realistic feasability study in the context of a future $e^{+}e^{-}$ Linear Collider such as TESLA, that takes into account all dominant physical backgrounds, the main radiative effects (initial and final state radiation and beamstrahlung) as well as detector and event reconstruction effects. A rough estimate of the loss of precision due to 6 fermion background processes was presented. The effect of the uncertainty on the top quark mass was also addressed. The $e^{+}e^{-} \rightarrow e^{+}e^{-}\gamma \gamma  \rightarrow e^{+}e^{-}$hadrons events, which may superimpose on physics events, were neglected. However, the impact of this background was studied for the $e^{+}e^{-} \rightarrow W^+W^-\nu \nu \rightarrow H \nu \nu $ process and found negligible \cite{bata_gaga}. Four channels were studied and the analysis repeated for several values of the Higgs boson mass ranging from 120 GeV/c$^{2}$ to 200 GeV/c$^{2}$. A collision energy of 800 GeV and an integrated luminosity of 1000 fb$^{-1}$ were assumed. The accuracies which are obtained account for the statistical uncertainty and for the systematic uncertainty arising from a limited knowledge of the background normalisation.

For Higgs boson masses under  $\approx $ 135 GeV/c$^{2}$, the main decay mode is  $H \rightarrow b\overline b$. Two channels were analysed - the hadronic and the semileptonic channels - and we presented the results for the Higgs mass range: 120 GeV/c$^{2}$ - 150 GeV/c$^{2}$. The measurement precision found for a mass of $120$ GeV/c$^{2}$ in the $H \rightarrow b\overline b$ mode is slightly worse than in~\cite{justemerino}, due to refinements of the present analysis which make it more realistic. The resolution degrades with increasing mass, due to the reduced event rate. In this decay mode, the ability to identify b-jets is of major importance.

For Higgs boson masses above $\approx $ 140 GeV/c$^{2}$, the $H \rightarrow W^{+}W^{-}$ decay mode yields higher precision than the $H \rightarrow b\overline b$ one as the number of signal events in the latter mode becomes too tiny. Two channels were analysed - the 2 like sign lepton plus 6 jet channel and the single lepton plus 8 jet channel - and we presented the results for the Higgs mass range: 130 GeV/c$^{2}$ - 200 GeV/c$^{2}$.

We showed that the 6 fermion background has a very modest influence on the measurement of the coupling, and that the present limited knowledge on the Standard Model input parameters has no significant influence on the predicted accuracy on $g_{ttH}$.

\begin{figure}[h]
\begin{center}
\vspace{-3.5cm}
\includegraphics[angle=90,width=15cm,height=21cm]{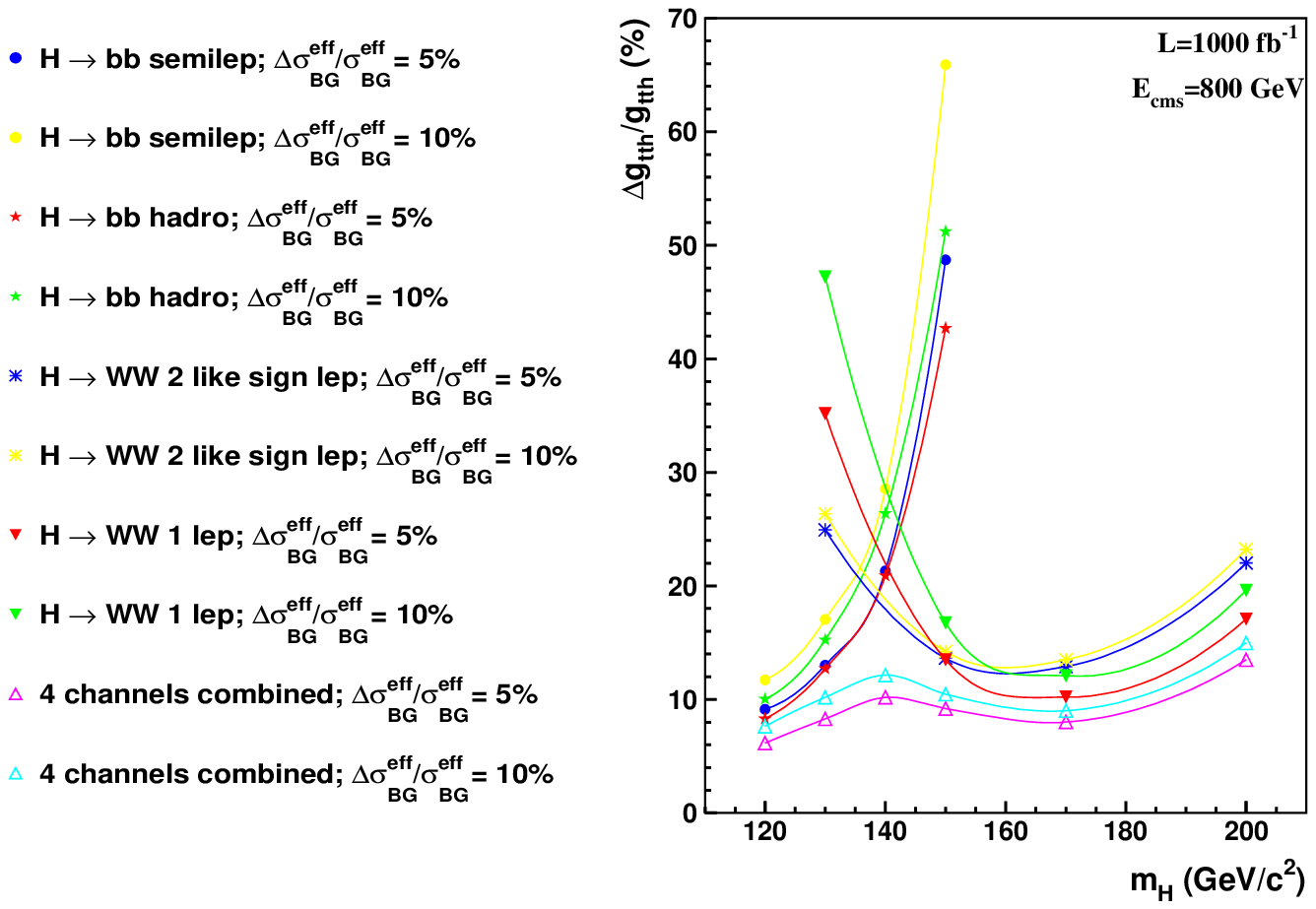}
\end{center}
\vspace{-2cm}
\caption{\it Expected relative uncertainty on the measurement of $g_{ttH}$ via the process $e^{+}e^{-} \rightarrow t\overline tH$ for various channels and their combination, for various Higgs boson masses and for two values of the relative uncertainty on the residual background normalisation.}
\label{plotresults}
\vspace{-3.5cm}
\end{figure}

As a final result, the four channels studied in this paper are combined\footnote{We combine the uncertainties found without inclusion of 6 fermion processes as the degradation due to them has only been roughly estimated and is anyway small.} to get the global precision and these results are shown in table~\ref{combi4channels} and on figure~\ref{plotresults}. 

The expected accuracy on $g_{ttH}$ is better than $\approx 10\%$ over most of the mass range (up to $M_{H} \approx $ 180 GeV/c$^{2}$), even if the knowledge of the background normalisation is only at the $10\%$ level. In the most favourable case ($M_{H}$= 120 GeV/c$^{2}$ and $\frac{\Delta \sigma _{BG}^{eff}}{\sigma _{BG}^{eff}}$ = 5$\%$), the accuracy is about 6$\%$. It is also good (8 to 9\%) for $M_{H} \approx$ 170 GeV/c$^{2}$, corresponding to the maximum of the branching ratio of the Higgs boson into pairs of W's. For the less favourable case ($M_{H}$= 200 GeV/c$^{2}$), the accuracy is however still better than 15$\%$ even if $\frac{\Delta \sigma _{BG}^{eff}}{\sigma _{BG}^{eff}}$ = 10$\%$.

We observe that the best resolutions are obtained for the lowest Higgs mass values, which are the most likely as predicted by precision measurements.

At the LHC, under some conditions which make the analysis more model-dependent, the expected accuracy on $g_{ttH}$ with 300 fb$^{-1}$ lies in the range $10\%-20\%$ for $M_{H} \in$ [100 GeV/c$^{2}$-200 GeV/c$^{2}$]  ~\cite{topyukawalhc2}. The precise measurements of the Higgs branching ratios available already at a low energy run of the Linear Collider will allow to remove the model-dependence of the LHC results \cite{lhclc}. The achievable accuracies are however not as good as those expected at a high energy run of the Linear Collider.

Finally, it should be stressed that there is certainly room for substantial improvement of the study exposed in this note. For instance, the b-tagging does not include all observables (e.g. vertex charge), more efficient jet reconstruction and particle flow algorithms could improve the reconstruction, other final states could be included, the reconstruction of $\tau$'s could enhance the number of signal events in leptonic channels and the analysis itself (selection criteria, neural network inputs, training and architecture) can also be optimized. Moreover, the analysis may be extended above $M_{H}$=200 GeV/c$^2$ in order to cover the full Higgs mass range allowed by precision measurements.

\begin{table}[h]
\centering
\begin{tabular}{||c||c||c|}
\hline
\hline
{ $M_{H}$ (GeV/c$^{2}$)} & {$\frac{\Delta \sigma _{BG}^{eff}}{\sigma _{BG}^{eff}}$} & {$\frac{\Delta g_{ttH}}{g_{ttH}}$}\\
\hline
\hline
{ 120} &{$5\%$}  & {$6.1\%$}\\
{    } &{$10\%$} & {$7.6\%$}\\
\hline
{ 130} &{$5\%$}  & {$8.3\%$}\\
{    } &{$10\%$} & {$10.2\%$}\\
\hline
{ 150} &{$5\%$}  & {$9.2\%$}\\
{    } &{$10\%$} & {$10.5\%$}\\
\hline
{ 170} &{$5\%$}  & {$8.0\%$}\\
{    } &{$10\%$} & {$9.0\%$}\\
\hline
{ 200} &{$5\%$}  & {$13.5\%$}\\
{    } &{$10\%$} & {$15.0\%$}\\
\hline
\hline
\end{tabular}
\caption{\it Expected relative uncertainty on the measurement of $g_{ttH}$ via the process $e^{+}e^{-} \rightarrow t\overline tH$ for the combination of the 4 channels studied, for various Higgs boson masses and for two values of the relative uncertainty on the residual background normalisation.} 
\label{combi4channels}
\end{table}

\section*{Acknowledgements}
\addcontentsline{toc}{section}{Acknowledgements}

I thank Marc Winter and Iouri Gornouchkine for valuable discussions.

\newpage
\addcontentsline{toc}{section}{List of figures}
\listoffigures

\newpage

\newpage

\newpage

\newpage

\newpage

\newpage

\newpage

\newpage

\newpage

\newpage

\newpage

\newpage

\newpage

\newpage

\newpage

\newpage

\newpage

\newpage

\newpage

\end{document}